\documentclass[10pt,preprint,twocolumn]{aastex7}
\newcommand{\roma}[1]{\uppercase\expandafter{\romannumeral#1}}
\newcommand{\speed}[1]{#1 km~s${}^{-1}$}

\newcommand{\nfig}[1]{Figure~\ref{#1}}

\usepackage{amsmath}
\usepackage{subfigure}
\usepackage{hyperref}

\usepackage{lineno}
\usepackage{color}
\usepackage{makecell}
\usepackage{graphicx}
\usepackage{natbib}
\usepackage{amssymb,txfonts}
\usepackage{multirow}
\usepackage{array}
\usepackage{sidecap}
\usepackage{hyperref}
\usepackage{epstopdf}
\usepackage{footnote}
\usepackage{tabularx}
\usepackage{booktabs}

\shorttitle{Failed Solar Filament Eruptions Associated with Magnetic Flux Emergence}
\shortauthors{Zhou et al.}
\received{March 31, 2025}
\revised{May 19, 2025}
\accepted{\today}

\begin{document}
\title{How Reconnection-Unfavored Magnetic Flux Emergence Suppresses Solar Filament Eruptions}
\correspondingauthor{Yuandeng Shen}

\author{Chengrui Zhou$^{1,5}$}
\noaffiliation{}
\affiliation{Yunnan Observatories, Chinese Academy of Sciences, Kunming 650216, China}
\affiliation{State Key Laboratory of Solar Activity and Space Weather, School of Aerospace, Harbin Institute of Technology, Shenzhen 518055, China}
\affiliation{Shenzhen Key Laboratory of Numerical Prediction for Space Storm, Harbin Institute of Technology, Shenzhen 518055, China}
\affiliation{School of Physics and Astronomy, Yunnan University, Kunming 650500, China}
\affiliation{University of Chinese Academy of Sciences, Beijing, 100049, China}
\email{zhouchengrui@ynao.ac.cn}  

\author[orcid=0000-0001-9493-4418]{Yuandeng Shen}
\affiliation{State Key Laboratory of Solar Activity and Space Weather, School of Aerospace, Harbin Institute of Technology, Shenzhen 518055, China}
\affiliation{Shenzhen Key Laboratory of Numerical Prediction for Space Storm, Harbin Institute of Technology, Shenzhen 518055, China}
\email[show]{ydshen@hit.edu.cn}

\author{Chun Xia}
\affiliation{School of Physics and Astronomy, Yunnan University, Kunming 650500, China}
\email{chun.xia@ynu.edu.cn}

\author{Hao Liang}
\affiliation{School of Physics and Astronomy, Yunnan University, Kunming 650500, China}
\email{824334107@qq.com}

\author{Zehao Tang}
\affiliation{Yunnan Observatories, Chinese Academy of Sciences, Kunming 650216, China}
\affiliation{University of Chinese Academy of Sciences, Beijing, 100049, China}
\email{tangzh@ynao.ac.cn}

\author{Dongxu Liu}
\affiliation{Yunnan Observatories, Chinese Academy of Sciences, Kunming 650216, China}
\affiliation{University of Chinese Academy of Sciences, Beijing, 100049, China}
\email{liudongxu@ynao.ac.cn}

\author{Surui Yao}
\affiliation{Yunnan Observatories, Chinese Academy of Sciences, Kunming 650216, China}
\affiliation{University of Chinese Academy of Sciences, Beijing, 100049, China}
\email{yaosurui@ynao.ac.cn}

\begin{abstract}
Magnetic flux emergence is traditionally considered a key trigger of solar filament eruptions; however, its role in suppressing filament eruptions remains less understood. Using multi-wavelength observations from the Solar Dynamics Observatory (SDO), this study investigates a unique case of {flux emergence} below a quiescent filament from January 3 to 5, 2016, where the newly emerging magnetic flux suppressed rather than promoted the eruption of the filament. It is found that the emerging magnetic bipole within the filament channel directly interacted and reconnected with the overlying filament magnetic field and produced a series of two-sided coronal jets along the filament axis. Instead of eruption, the filament kept stable but broke into two segments at the reconnection site. Further magnetic cancellation or recession of the emerged bipole allowed the filament to recover its original structure. Our analysis results revealed that the flux emergence suppressed the filament eruption by reducing the upward net force. The formation and evolution of the filament fine structures (such as filament threads) are closely linked to the reconnection processes between the emerging bipole and the filament’s horizontal magnetic field. This study provides direct observational evidence for accounting for the stabilization of solar filaments driven by {flux emergence}, offering new insights into magnetic emergence's dual role in triggering and suppressing solar eruptions.
\end{abstract}

\section{Introduction}\label{intro}
Filaments are observed as dark elongated absorption features in H$\alpha$ images on the solar disk; they are also referred to as prominences when viewed above the disk limb. Previous studies indicated that filaments are cool, dense plasma supported by the magnetic field in the hot, tenuous corona, and their violent eruptions are often accompanied by flares and coronal mass ejections (CMEs) that might lead to disastrous space weather in the near-earth space environment \citep[e.g.,][]{2011LRSP....8....1C,2013ScChD..56.1091W}. Generally, a filament can exist for several days or even a few months, but sometimes, it can lose its stability and erupt violently into interplanetary space. According to previous studies, many physical reasons are proposed to {cause filaments to be unstable and erupt}, including the emergence and cancellation of magnetic fluxes~\citep[e.g.,][]{1995JGR...100.3355F,chen00,2012A&A...537A..62A,2024ApJ...964..125S,shen19b,2017ApJ...851..101S}, ideal magnetohydrodynamic (MHD) instabilities such as kink instability and torus instability~\citep[e.g.,][]{ji03,kumar11,joshi13,joshi14,xuh20, Tor05,kliem06, Liuy08, Liu12}, and external disturbances including but not limited to nearby flares~\citep[e.g.,][]{2020ApJ...892...79S,2021ApJ...923...45Z,hou20}, CMEs~\citep[e.g.,][]{shen12b,yang12,jiang11, Lynch13,tor11,2023ApJ...943...62Y}, coronal jets~\citep[e.g.,][]{shen12a}, and MHD waves~\citep{2014ApJ...786..151S, 2025RAA....25a5006L}.

Observational and theoretical studies have demonstrated that the orientation of newly emerged magnetic flux plays a fundamental role in filament eruptions. When this flux is aligned to facilitate reconnection with the filament's confined field, it can destabilize the magnetic system and trigger the eruption of the filament. For example, \citet{1995JGR...100.3355F} showed that a reconnection-favored magnetic flux emergence destabilizes the magnetic structure of a filament and leads to the eruption. This mechanism was further confirmed by \citet{chen00} using the method of numerical simulation, which demonstrated that emerging fluxes near pre-existing current-carrying structures could weaken the overlying field of the latter and, therefore, initiate the eruption of the magnetic system. \citet{2004A&A...426.1047A} and \citet{2012A&A...537A..62A} conducted 3D MHD simulations to study the interactions between emerging twisted flux tubes and pre-existing horizontal coronal fields. They found that reconnection between emerging bipolar loops and the existing coronal field can produce dynamic solar events, such as coronal jets and filament eruptions. Previous studies highlighted that reconnection-favored emergence can serve as a dominant eruption trigger \citep{2006A&A...459..927D,2021ApJ...912...97H}. However, magnetic flux emergence does not always lead to filament eruptions. On the contrary, some magnetic flux emergence can suppress the eruption of filaments. The dual role of magnetic flux {emergence}—both triggering and suppressing filament eruptions—remains unresolved, necessitating further studies to fully understand its impact on filament dynamics.

While magnetic flux emergence is widely recognized as a key trigger for solar eruptions, its role in suppressing filament eruptions remains less understood. This suppression is associated with the reconnection-unfavored emergence of magnetic flux, where the newly emerged magnetic flux is oriented parallel to the background magnetic field. In this case, it is thought that the emerging magnetic flux cannot easily reconnect with the pre-existing filament above. For instance, \citet{chen00} found that the newly emerging magnetic flux can stabilize the overlying filament by enhancing the magnetic tension force of the filament's confining magnetic field. Recent 3D MHD simulations further indicate that while magnetic flux emergence generally triggers solar eruptions, the actual occurrence of an eruption also depends on factors such as the twist of the emerging flux and its orientation relative to the background magnetic flux rope \citep{2024ApJ...962..149T}. Despite these advancements, the criteria distinguishing eruption-triggering from eruption-suppressing emergence, particularly the role of flux emergence in stabilizing filaments and its influence on their 3D magnetic configuration, remain unquantified in observational studies.

This study investigates the evolution of a filament caused by the emergence of a magnetic bipole within the filament channel. The high spatiotemporal resolution and multiwavelength observations clearly showed the detailed interaction and reconnection between the emerging bipole and the pre-existing filament, during which a series of successive two-sided jets are observed originating from the reconnection site, and the filament first {broke} into two segments and then recovered gradually during the cancellation or recession of the bipole. We aim to study how the {flux emergence with a reconnection-unfavorable direction within the filament channel} suppresses the filament eruption and what kind of characteristic signal can be observed for diagnosing the changing of the filament magnetic structure. The observational data and instruments used in this study are introduced in Section ~\ref{sec:data}; {the} main analysis results are presented in Section~\ref{sec:result}; Interpretations and discussions are given in Section~\ref{sec:interp}, and the last section~\ref{sec:summary} is the summary of our study.

\section{Observations and Methods}\label{sec:data}
The event was observed by the Atmospheric Imaging Assembly \citep[AIA;][]{lemen12} and Helioseismic and the Magnetic Imager \citep[HMI;][]{Schou2012} onboard the Solar Dynamics Observatory \citep[{\em SDO};][]{pesnell12}. AIA captures solar plasma across various temperatures using multiple filters, including two ultraviolet (UV) and seven extreme ultraviolet (EUV) wavelengths. For this study, we primarily utilized AIA channels at 171~\AA~(0.6 MK), 193~\AA~(1.6 MK), and 304~\AA~(50,000 K). These full-disk, multi-wavelength images have a pixel size of 0. $^{\prime \prime}$6 and a cadence of 12 seconds. HMI provides full-disk magnetograms with a pixel size of 0.$^{\prime \prime}$5, capturing line-of-sight (LOS) photospheric magnetograms every 45 seconds and photospheric vector magnetograms every 12 minutes. This favorable viewing angle and orientation allowed us to observe the detailed evolution of the filament's fine structure during the magnetic flux emergence. To model the 3D coronal magnetic field surrounding the filament, we performed a Potential Field Source Surface (PFSS) extrapolation using HMI photospheric vector magnetic fields as the lower boundary input.

The evolution of magnetic helicity in the solar atmosphere is governed by two primary mechanisms: (1) the emergence of pre-twisted magnetic flux through the photosphere (emergence term, $\mathrm{d}H_{\mathrm{e}}/\mathrm{d}t$), and (2) the generation through photospheric shearing motions (shear term, $\mathrm{d}H_{\mathrm{s}}/\mathrm{d}t$). The total helicity flux over a closed surface $S$ can be expressed as \citet{1984JFM...147..133B}:

\begin{equation}\label{eq:total_helicity}
\left.\frac{\mathrm{d}H}{\mathrm{d}t}\right|_{S} = 
2\underbrace{\int_{S}(\vec{A}_{\mathrm{p}}\cdot \vec{B}_{\mathrm{t}})V_{\mathrm{n}}\,\mathrm{d}S}_{\text{Emergence Term }(\mathrm{d}H_{\mathrm{e}}/\mathrm{d}t)} 
- 2\underbrace{\int_{S}(\vec{A}_{\mathrm{p}}\cdot \vec{V}_{\mathrm{t}})B_{\mathrm{n}}\,\mathrm{d}S}_{\text{Shear Term }(\mathrm{d}H_{\mathrm{s}}/\mathrm{d}t)},
\end{equation}
where $\vec{A}_{\mathrm{p}}$ is the vector potential of the potential field $\vec{B}_{\mathrm{p}}$, $\vec{B}_{\mathrm{t}}$ and $B_{\mathrm{n}}$ denote the tangential and normal magnetic field components, and $\vec{V}_{\mathrm{t}}$, $V_{\mathrm{n}}$ represent the tangential and normal components of plasma velocity $\vec{V}$.

A critical advancement came from \citet{2003SoPh..215..203D}, who introduced the concept of flux transport velocity $\vec{u}$ to describe the apparent footpoint motion:

\begin{equation}\label{eq:flux_velocity}
\vec{u} = \vec{V}_{\mathrm{t}} - \frac{V_{\mathrm{n}}}{B_{\mathrm{n}}}\vec{B}_{\mathrm{t}},
\end{equation}
where the second term accounts for flux tube distortion caused by vertical flows. Building on this framework, \citet{2005A&A...439.1191P} reformulated the helicity flux using a pairwise interaction approach:

\begin{equation}\label{eq:pariat_helicity}
\left.\frac{\mathrm{d}H}{\mathrm{d}t}\right|_{S} = \frac{-1}{2\pi} \iint\limits_{S\times S} 
\frac{[(\vec{x}-\vec{x'})\times(\vec{u}-\vec{u'})]_{\mathrm{n}}}{|\vec{x}-\vec{x'}|^2} 
B_{\mathrm{n}}(\vec{x})B_{\mathrm{n}}(\vec{x'})\,\mathrm{d}S\mathrm{d}S',
\end{equation}
where $\vec{x}$ and $\vec{x'}$ denote footpoint positions of interacting flux tubes.

Subsequent studies \citep{2012ApJ...761..105L} emphasized the need to exclude field-aligned plasma flows ($\vec{V}_{\parallel}$) that do not contribute to helicity transport. This leads to a revised expression for $\vec{u}$:

\begin{equation}\label{eq:revised_velocity}
\vec{u} = \vec{V}_{\perp\mathrm{t}} - \frac{V_{\perp\mathrm{n}}}{B_{\mathrm{n}}}\vec{B}_{\mathrm{t}},
\end{equation}
where $\vec{V}_{\perp} \equiv \vec{V} - (\vec{V}\cdot\vec{B}/B^2)\vec{B}$ represents the flow component perpendicular to the magnetic field. Correspondingly, the emergence and shear terms become:

\begin{equation}
\begin{split}
\frac{{\rm d}H_{\rm e}}{\rm dt}=\frac{1}{2 \pi}\int_{S}\int_{S^{\prime}} {\frac{\vec{x}-\vec{x^{\prime}}}{|\vec{x}-\vec{x{^{\prime}}}|^2}}{\rm d}S{\rm d}S^{\prime}
\times\\
\\
[\vec{B}_{\rm t}(\vec{x})V_{\bot {\rm n}}(\vec{x})B_{\rm n}(\vec{x^{\prime}})-\vec{B}_{\rm t}(\vec{x^{\prime}})V_{\bot {\rm n}}(\vec{x^{\prime}})B_{\rm n}(\vec{x})],
\end{split}
\label{eq:1}
\end{equation}

\begin{equation}
\begin{split}
\frac{{\rm d}H_{\rm s}}{\rm dt}=\frac{-1}{2 \pi}\int_{S}\int_{S^{\prime}} {\frac{\vec{x}-\vec{x^{\prime}}}{|\vec{x}-\vec{x{^{\prime}}}|^2}}{\rm d}S{\rm d}S^{\prime}\times\\
\\
[(\vec{V}_{\bot {\rm t}}(\vec{x})-\vec{V}_{\bot {\rm t}}(\vec{x^{\prime}}))B_{\rm n}(\vec{x})B_{\rm n}({\vec{x^{\prime}})}],
\end{split}
\label{eq:2}
\end{equation}

Here, $\vec{B}_{{\rm n}}$ and $B_{\rm t}$ are the normal and tangential magnetic fields, respectively, and $\vec{V}_{\rm t}$ and $\vec{V}_{\rm n}$ are the tangential and normal components of the plasma velocity $\vec{V}$, $\vec{x}$ and $\vec{x^\prime}$ denote the photospheric footpoints of magnetic flux tubes, which can be obtained using the Differential Affine Velocity Estimator for Vector Magnetograms (DAVE4VM) flow tracking method \footnote{DAVE-DAVE4VM flow tracking codes can be obtained from http://ccmc.gsfc.nasa.gov/lwsrepository/index.php} flow tracking~\citep{2008ApJ...683.1134S}.

The surface density of helicity flux $G_\theta$ satisfies:

\begin{equation}\label{eq:helicity_density}
\int_{S} G_{\theta} \,\mathrm{d}S = \left.\frac{\mathrm{d}H}{\mathrm{d}t}\right|_{S} = \left.\frac{\mathrm{d}H_{\mathrm{e}}}{\mathrm{d}t}\right|_{S} + \left.\frac{\mathrm{d}H_{\mathrm{s}}}{\mathrm{d}t}\right|_{S}.
\end{equation}

However, as helicity is inherently non-local \citep{2005A&A...439.1191P}, $G_\theta$ only acquires physical meaning when considering complete flux tube pairs. This led to the improved definition by \citet{2005A&A...439.1191P}:

\begin{equation}\label{eq:improved_density}
G_\phi (\vec{x}_{\mathrm{a}\pm}) = \frac{1}{2} \left[ G_\theta(\vec{x}_{\mathrm{a}\pm}) + G_\theta(\vec{x}_{\mathrm{a}\mp}) \left|\frac{B_{\mathrm{n}}(\vec{x}_{\mathrm{a}\pm})}{B_{\mathrm{n}}(\vec{x}_{\mathrm{a}\mp})}\right| \right],
\end{equation}
where $\vec{x}_{\mathrm{a}\pm}$ denote conjugate footpoints of a closed flux tube. This formulation ensures proper helicity redistribution between connected magnetic elements \citep{2014SoPh..289..107D}.
 
\section{Observational Results}\label{sec:result}
\begin{figure*}
\centering
\includegraphics[width=0.9\textwidth]{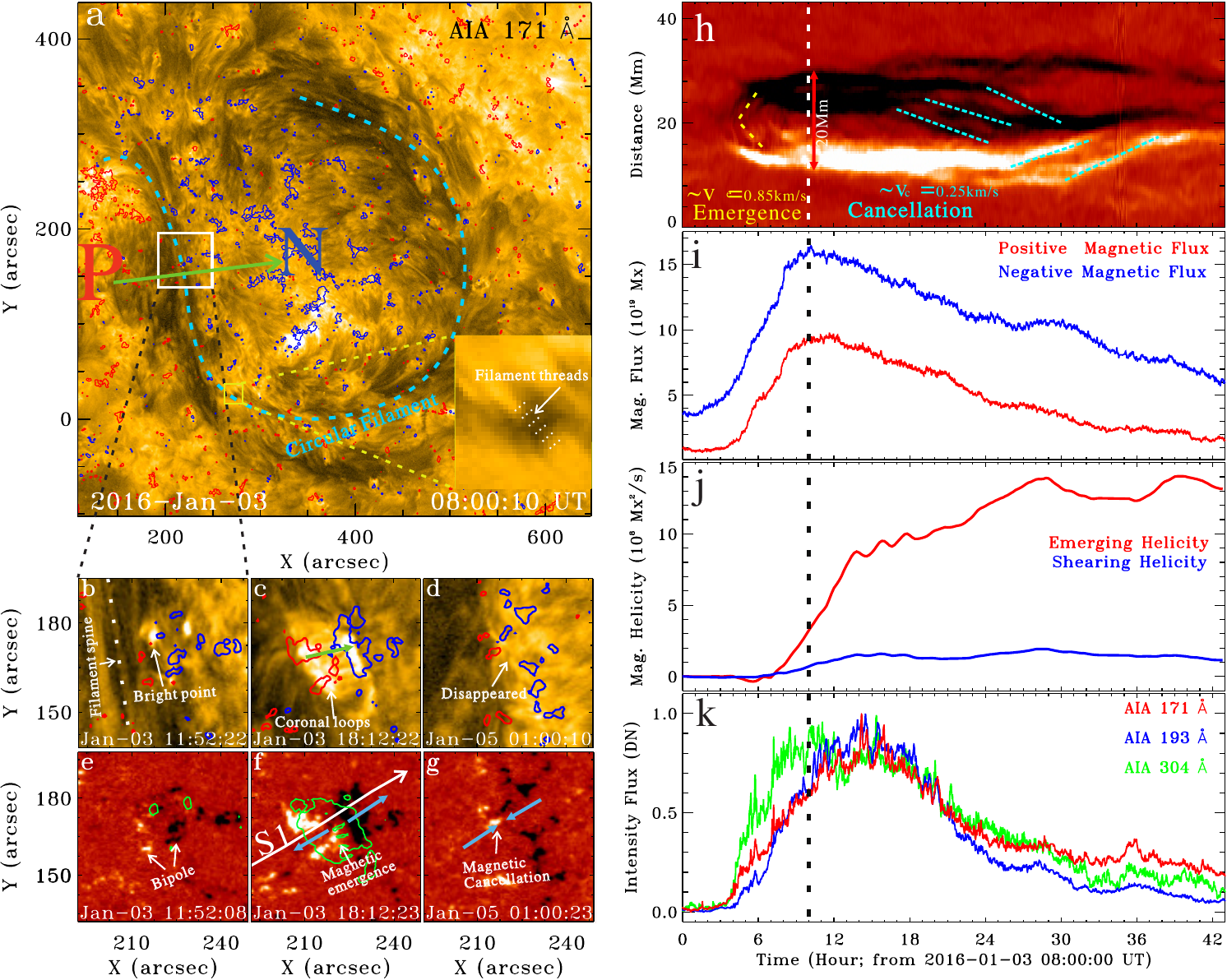}
\caption{AIA 171\AA~images and HMI line-of-sight (LOS) magnetograms illustrate the magnetic emergence and cancellation process of a bipole. Panel (a) provides an overview of the circular filament structure, the blue dashed lines denote the spine of the circular filament, and the LOS magnetogram at 08:00:45 UT is overlaid at $\pm100$ G, while the red (blue) color indicates positive (negative) polarity, {while the white box defines the field of view (FOV) for panels (b)-(g)}, and the region for AIA light curve/magnetic analysis in panels (i)-{(k)}. The green arrow in panel (a) indicates the overlying magnetic field orientation of the filament channel (perpendicular to the filament spine). Panels (b)-(d) exhibit the time sequence of the bipole region in AIA 171~\AA~observations, with corresponding magnetograms overlaid. Panels (e)-(g) show the co-temporal LOS magnetogram corresponding to panels (b)-(d), the green contours are the bright features in the co-temporal AIA 171 \AA~observations. Panel (h) shows the time–distance diagram obtained by S1 in \nfig{fig1} (f), where the yellow dashed lines denote expansion during the emergence phase, while the blue dashed lines show that the bipole got close to each other. Panels (i)-(k) show the temporal evolutions of magnetic flux, magnetic helicity, and the AIA light curves during the magnetic emergence and cancellation in the magnetic bipolar region. {An animation of the AIA 304 \AA\ (left), {AIA 171} \AA\ (middle), and AIA 193 \AA\ (right) images of this event is available in the online Journal. The animated images include both the wide scale and zoom scale of the event.}}
\label{fig1}
\end{figure*}

On January 03, 2016, a circular quiescent filament located in the solar northern hemisphere was observed by the SDO as indicated by the blue dashed line in the AIA 171 \AA\ image at 08:00:10 UT (\nfig{fig1} (a)). One can find that the eastern portion of the filament has an inversed S-shape, which means that the magnetic twist of the filament should be negative, as an inverted S-shape filament in the northern hemisphere is typically dominated by negative {chirality}~\citep{1998SoPh..182..107M, 2020RAA....20..166C}. We noted some dark thread-like features in the yellow box region in \nfig{fig1} (a), which might represent a magnetic dip that trapped more cold filament plasma than other parts~\citep[see also the white dashed lines in the inset;][]{2003ApJ...593.1187K,2021ApJ...920..131G}. {It should be noted that filament threads exhibit a near-parallel orientation to the filament polarity inversion line, shows the condition of strong shearing for the sheared arcade model or the flux rope model with weak twist. This alignment may suggest a transition between sheared arcades and flux ropes, with a twist number close to 1~\citep{2020RAA....20..166C}}. The white box in \nfig{fig1} (a) indicates the region where the magnetic bipole emerged. The temporal evolution of the emerging bipole is displayed in \nfig{fig1} (b) -- (g) using the AIA 171 \AA\ images and the HMI magnetograms. \nfig{fig1} (b) and (e) show the early stage of the emerging bipole at 11:52 UT, in which one can see that the emerging positive and negative magnetic polarities in the HMI magnetogram well correspond to the two bright points in the AIA 171 \AA\ image.  About six hours later, at 18:12 UT, while the positive and negative magnetic polarities in the HMI magnetogram grew a lot in size, the two bright points also evolved into a large group of bright coronal loops in the AIA 171 \AA\ image (see \nfig{fig1} (c) and (f)). On January 05, 2016, both the coronal loops and opposite magnetic polarities disappeared (see \nfig{fig1} (d) and (g)). The temporal and spatial correlations between the dynamics of the magnetic bipole and the corresponding coronal loops are consistent with the evolution of magnetic emergence ~\citep{2004A&A...426.1047A,2012A&A...537A..62A}. By comparing the location and magnetic polarity of the emerging bipole with those of the background magnetic field, it reveals that the emerging bipole belongs to the so-called {flux emergence with a reconnection-unfavorable direction within the filament channel} \citep{1995JGR...100.3355F, chen00}. 
 
\begin{figure*} 
\centering
\includegraphics[width=0.9\textwidth]{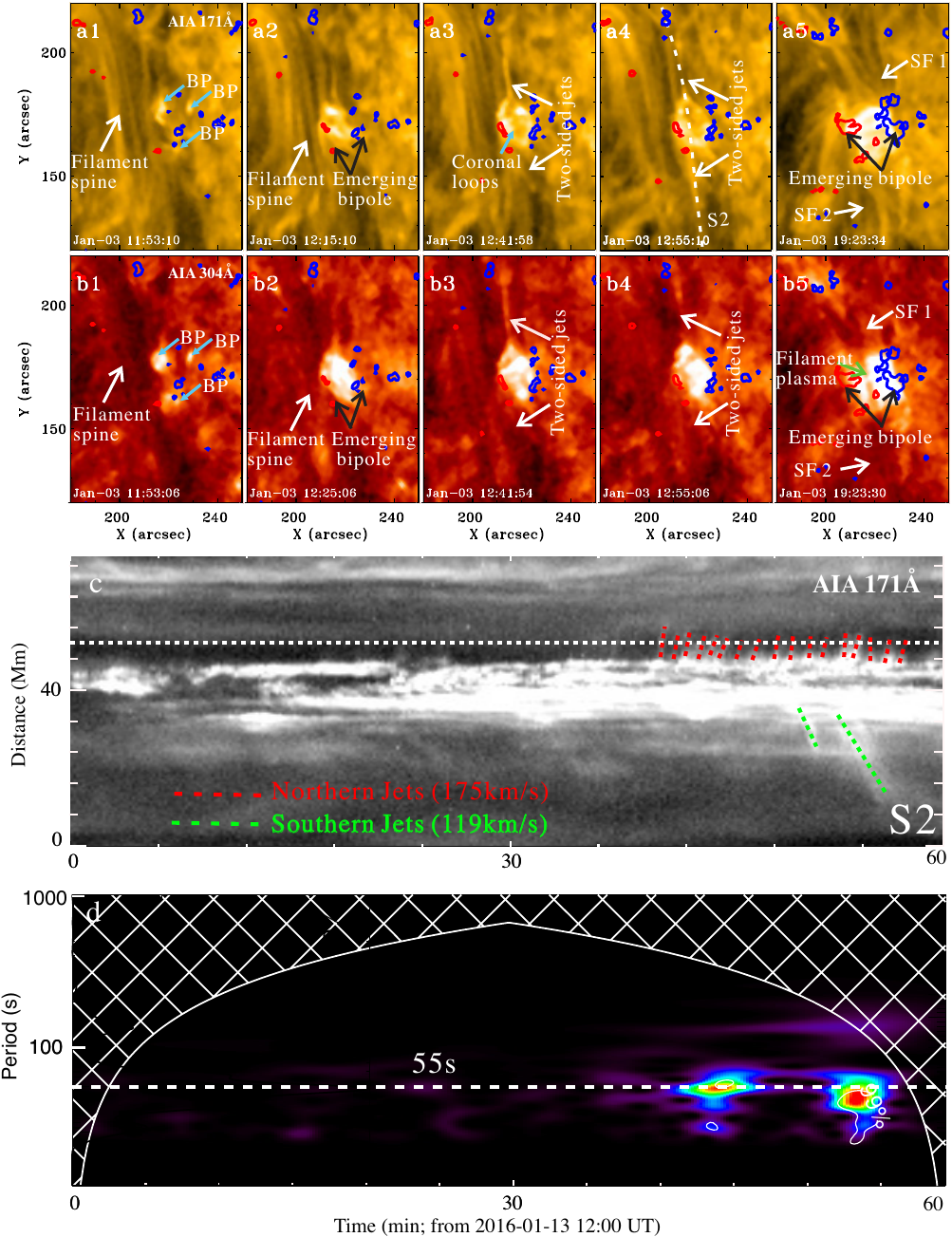}
\caption{\nfig{fig2} (a1)-(b5) show the time sequence of the two-sided jets in the emergence phase of the bipolar region in AIA 171 \AA~and AIA 304 \AA~images, in each panel, the co-temporal LOS magnetogram is overlaid at $\pm100$ G, the red (blue) color indicates positive (negative) polarity. \nfig{fig2} (c) presents the time-distance diagrams derived from the white dashed line S2 in panel (a4), and the red (green) dashed lines in pane (c) show the trajectory of the northern (southern) jets, while the white dashed line denotes the data for wavelet analysis. \nfig{fig2} (d) exhibits the wavelet analysis of the 171~\AA~emission intensity of the jets. The white dashed line shows the period of the jets, about 55 seconds. {An animation of the AIA 304 \AA\ (left), {AIA 171} \AA\ (middle), and AIA 193 \AA\ (right) images of this event is available in the online Journal. The animated images include both the wide scale and zoom scale of the event.}
\label{fig2}}
\end{figure*}

\nfig{fig1} (h) -- (k) shows the kinematic evolution of the magnetic bipole and the variations of the associated fluxes from 08:00 UT on January 03 to 03:00 UT on January 05, its animation is available in the online journal. \nfig{fig1} (h) is a time-distance diagram made from the HMI magnetogram along the path S1 in \nfig{fig1} (f), from which one can see that the positive and negative polarities of the emerging magnetic bipole firstly experienced a fast separation in distance accompanying by a simultaneous rapid growing in size. The separation distance reached a maximum value of about 20 Mm at about 18:00 UT on January 03, during which the average emergence speed was about \speed{0.85} (see the yellow dashed lines). Then, the opposite magnetic polarities of the emerged magnetic bipole got close to each other, and the approaching speed of the polarities was about  \speed{0.25} during the later stage (see the blue dashed lines). In this paper, the term cancellation encompasses both potential processes—flux submergence and reconnection-driven cancellation, with submergence likely playing a primary role. We further investigated the variations of magnetic fluxes, magnetic helicity, and EUV intensity within the white box in \nfig{fig1} (a), and the results are plotted in \nfig{fig1} (i) -- (k), respectively. The curves of the magnetic fluxes indicate that the positive and negative magnetic fluxes first increased rapidly and peaked at about 18:00 UT on January 03, then both the positive and negative fluxes underwent a 31-hour-long decrease phase. The vertical dashed line in the figure indicates the watershed between the increase and decrease phases of the magnetic fluxes. \nfig{fig1} (j) shows the variation of the {local} magnetic helicity of the magnetic bipole by equation~\ref{eq:1} and equation~\ref{eq:2}. One can see that both the emerging and shearing magnetic helicities of the emerging bipole are positive, which is opposite to the negative magnetic chirality of the filament channel. While the emerging helicity monotonically increases at a large rate of increase during the entire lifetime of the bipole, the shearing helicity increases gradually at a small rate of increase, and its value is far less than the emerging helicity. The intensity lightcurves generally show a similar trend as the magnetic fluxes (see \nfig{fig1} (k)), but the peak times of the EUV lightcurves are delayed with respect to the magnetic fluxes by 4- 5 hours. It is noted that during the emerging stage of the magnetic fluxes, there are many small but prominent peaks on the lightcurves. Here, the main component of the rising lightcurves came from the hot emerging loops, while the small peaks were due to the reconnection of the emerging loops and the overlying filament channel. This means that the heating of the corona by magnetic flux emergence also has two ways, i.e., the hot emerging loops and their reconnection with the ambient magnetic field~\citep{1988ApJ...330..474P, 2024NatAs...8..697B,2024NatAs...8..706L}.

\nfig{fig2} shows the detailed interaction and reconnection between the emerging magnetic bipole and the overlying filament. The top two rows, respectively, display the AIA 171 \AA\ and 304 \AA\ time series images overlaid with the contours of the corresponding positive (red) and negative (blue) polarities. At 11:53 UT on January 03, the filament can be observed clearly in the AIA images,  and several bright points (or small loops) were on the west side of the filament (see the white and blue arrows in \nfig{fig1} (a1) and (b1)). {As the magnetic bipole emerged}, those bright points turned into {a cluster of low coronal loops, and started} to interact and reconnect with the filament, which produced several two-sided jets (see \nfig{fig2} (a2) -- (a4) and (b2) -- (b4)). After the reconnection process, the filament was broken into two segments from the reconnection site, with the south footpoint of the north segment (SF1) and the north footpoint of the south segment (SF2) rooted in the negative and positive polarities of the emerged magnetic bipole, respectively {(\nfig{fig2} (a5) and (b5))}. {It should be noted that 304~\AA~observation (see in \nfig{fig2} (b5)), the continuous filamentary structure persists during the peak of bipole emergence. This residual filament plasma indicates that while magnetic reconnection had eroded parts of the filament’s magnetic configuration, the core filament channel remained.} Assuming that the emerged coronal loops had a semicircular shape, thus, the {average} height of the filament can be roughly estimated from the separation distance of the bipole, yielding approximately 20 Mm. During the emerging phase of the magnetic bipole from 12:38 to 12:55 UT, there were at least 17 two-sided jets that could be directly observed ({the movie2.mpg available in the journal}). \nfig{fig2} (c) displays the time-distance diagram along the trajectory of the two-sided jets as shown by the white dashed line  S2 in \nfig{fig2} (a4), based on which we measured that the average speed of the northern (southern) jets is about \speed{171 (119)}. In addition, the northern jets exhibit a periodic behavior with a period of about 55 seconds, as revealed by the wavelet analysis of the intensity variation along the white dashed line in \nfig{fig2} (c). {This suggests that the energy release of the emerging magnetic bipole was in a stepwise magnetic reconnection.}

\begin{figure*} 
\centering
\includegraphics[width=0.85\textwidth]{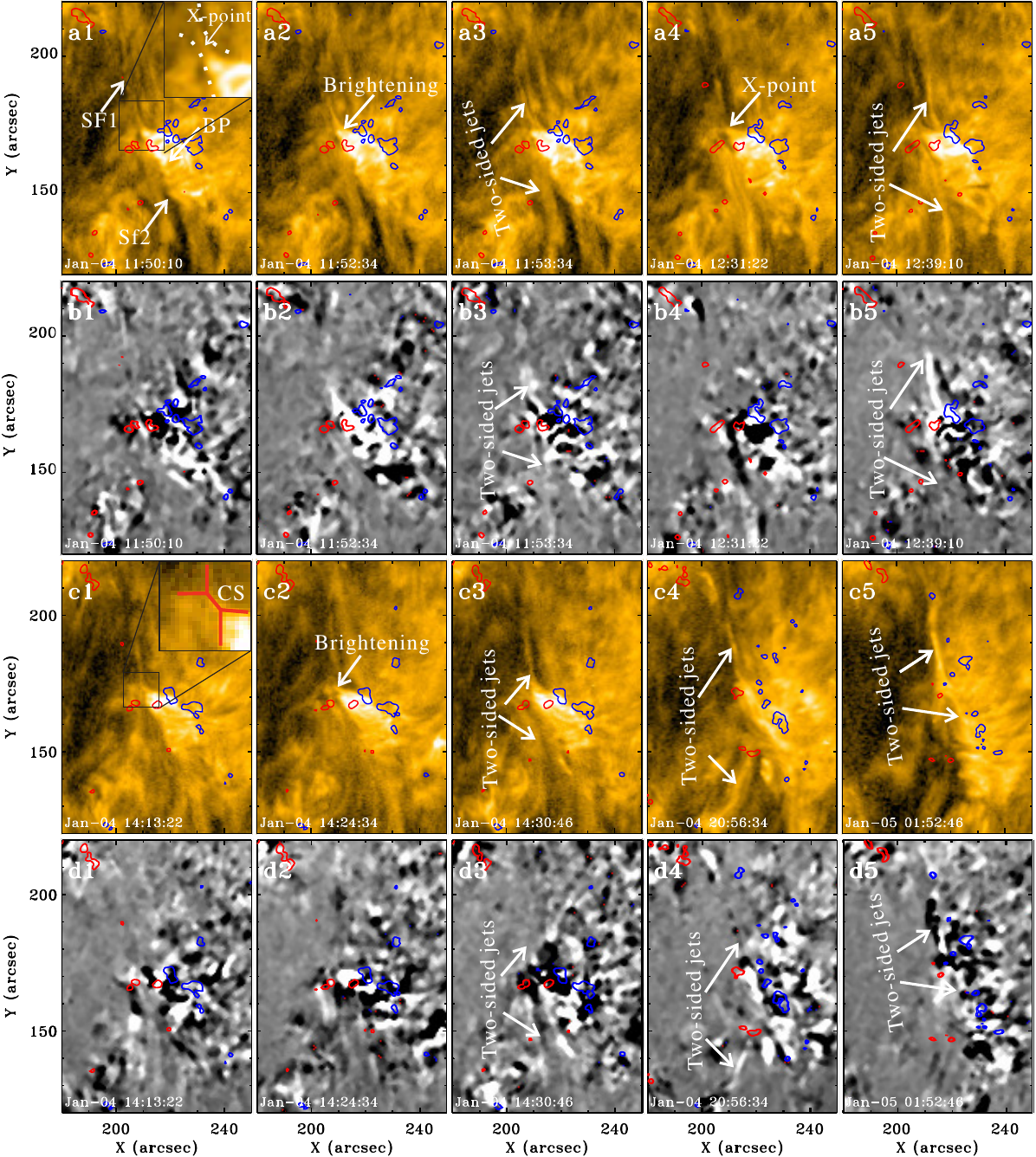}
\caption{Panels show the time sequence of some of the two-sided jets in the cancellation phase of the bipolar region in AIA 171 \AA~images and the co-temporal AIA 171~\AA~running difference images. In each panel, the co-temporal LOS magnetogram is overlaid at $\pm100$ G, and the red (blue) color indicates positive (negative) polarity.}
\label{fig3}
\end{figure*}

\nfig{fig3} displays the AIA 171 \AA\ direct and running-difference images to show the detailed evolution of the two filament segments (SF1 and SF2) during the magnetic cancellation phase. It can be observed that the south (north) footpoint of SF1 (SF2) {is} rooted in the emerged negative (positive) magnetic polarity (see the white dashed lines in \nfig{fig3} (a)), the crossed magnetic field lines of the two filament segments formed a reconnection-favored X-shaped magnetic configuration (see the insert in \nfig{fig3} (a1)). The crossed magnetic field {of filament} created a current sheet (CS) structure as the two opposite magnetic polarities approached each other for flux cancellation. As the positive and negative magnetic polarities got closer, magnetic reconnection between SF1 and SF2 caused their coalescence to form a long filament {channel} as the original one before {it broken} (see the insert image in \nfig{fig3} (c1)). The magnetic reconnection between SF1 and SF2 also produced brightening at the reconnection site (see the white arrows in \nfig{fig3} (a2) and (c2)) and several two-sided jets along the filament (see the white paired arrows in \nfig{fig3}). It is noted that the formation of the X-shaped magnetic configuration between SF1 and SF2 repeated a few times, and the brightening also appeared a few times at the X-shaped site. This might indicate that the intermittent approaching of the opposite emerged magnetic polarities during the cancellation stage. Similar physical processes in the formation of long filament {channel} or two-sided jets through reconnection between two short ones in association with magnetic flux cancellations were reported in many studies \citep[e.g.,][]{tian2017, 2016ApJ...816...41Y, 2017ApJ...840L..23X, zhengr17, yangbo19, 2016ApJ...818L..27C, 2024ApJ...975L...5Y}. In our case,  it is interesting that we observed the filament's complete breaking and recovery processes, in which the breaking was due to the reconnection between the filament{-supporting magnetic field} and the emerging magnetic bipole in the filament channel, while the recovery was due to the reconnection between the two broken filament segments.

\begin{figure*}
\centering
\includegraphics[width=0.9\textwidth]{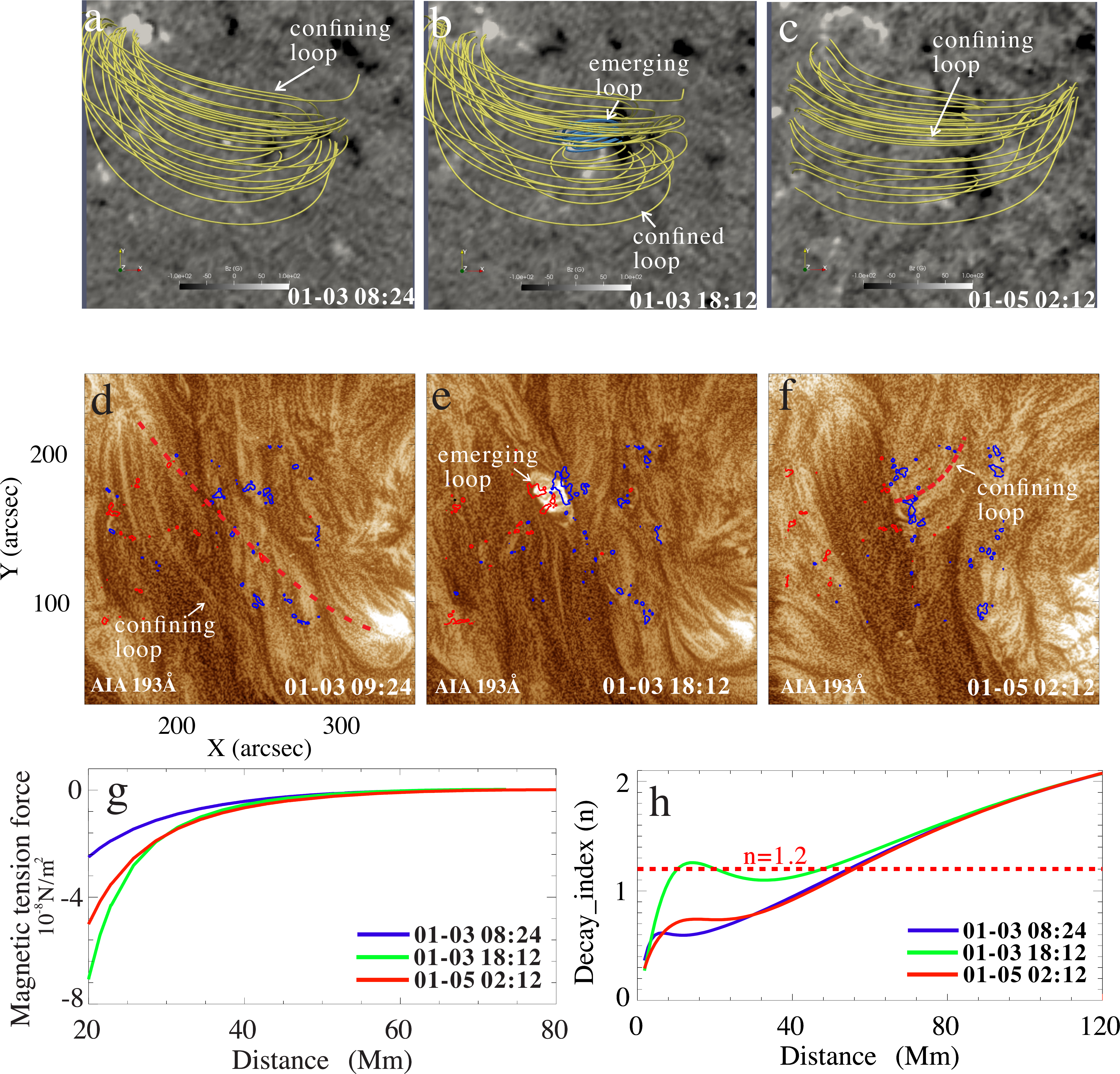}
\caption{{Panels (a)-(c) show some arcades that overlie the filament, as extrapolated by the PFSS calculation.} Panels (d)- (f) exhibit the MGN enhanced cotemporal AIA 193~\AA~images, the red dashed lines denote the {sample confining loops over} the filament, the co-temporal LOS magnetogram is overlaid at $\pm100$ G, the red (blue) color indicates positive (negative) polarity. Panels (g) and (h): the calculated magnetic tension force and decay index from the 3D PFSS extrapolated magnetic field, while the red dashed line in panel (h) labels the line n=1.2.}
\label{fig4}
\end{figure*}

\begin{figure*}
\centering
\includegraphics[width=0.85\textwidth]{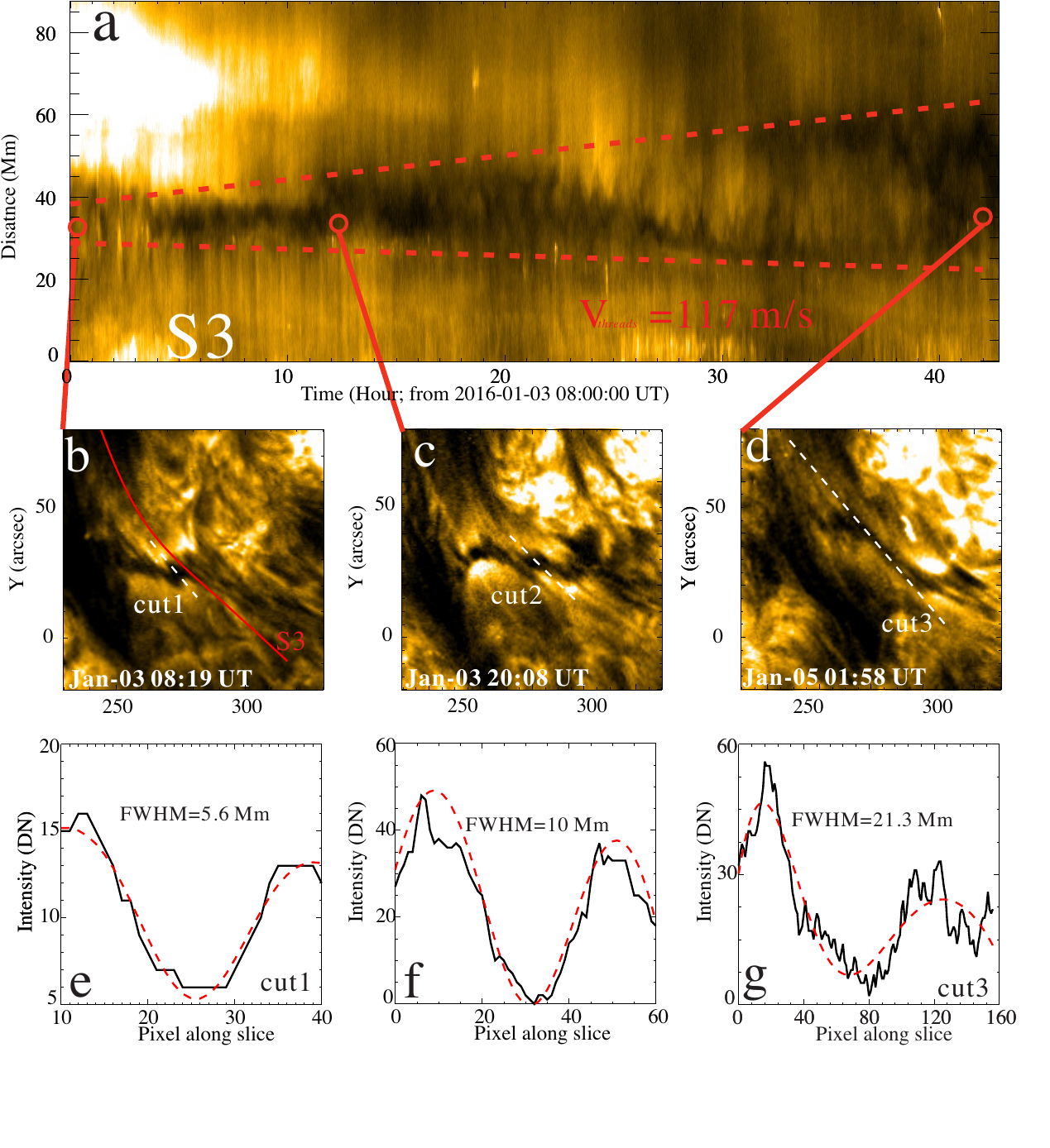}
\caption{\nfig{fig5} (a) shows the time-distance image from the red line S3 in panel (b). {Panels (b)-(d) show a zoomed FOV outlined by the yellow box in \nfig{fig1} (a), which} shows the filament threads, and the white dashed lines in each panel denote the line cut 1--3. Panel (k) shows the intensity curve along the line cut 1--3, and the red dashed lines denote the Gaussian fitting curves. {An animation of the AIA 304 \AA\ (left), {AIA 171} \AA\ (middle), and AIA 193 \AA\ (right) images of this event is available in the online Journal.}
\label{fig5}}
\end{figure*}

Using the HMI photospheric vector magnetograms from the SDO, we performed a Potential Field Source Surface (PFSS) extrapolation to reconstruct the three-dimensional coronal magnetic field surrounding the filament. {The PFSS method is well-suited for modeling quiet-Sun regions dominated by potential magnetic fields. However, it has inherent limitations in resolving non-potential components, such as electric currents and sheared magnetic fields. In our study, we adopted the PFSS approach mainly because of the relatively simple magnetic configuration observed in the {the confining loops of the filament channel}. Specifically, we assumed that non-potential effects played a secondary role at altitudes above 20 Mm, where the magnetic field lines closely approximate a potential field configuration.} Three key moments were selected for analysis: during the pre-emergence phase at 08:24 UT on January 3, the late emergence phase at 18:12 UT on January 3, and the late cancellation phase at 02:12 UT on January 5. As shown in Figure~\ref{fig4}(a)--(c), the extrapolated emerging and confining loops of the filament {channel} closely resemble those observed in the {Multi-Scale Gaussian Normalization (MGN) enhanced} AIA 193 \AA\ images in \nfig{fig4} {(d)}--(f) \citep{2014SoPh..289.2945M}. It can be observed that the confining loops of the filament {channel} experienced a gradual reorientation from a southeastward direction to an east-west alignment, which indicates a progressive reduction in the filament's {twist}~\citep{1998SoPh..182..107M}. Figure~\ref{fig4}(g) displays the calculated magnetic tension force ($\mathbf{T}_B = \frac{1}{\mu_0} (\mathbf{B} \cdot \nabla) \mathbf{B}$) within the altitude range of 20 to 90 Mm above the emerging bipolar region. The magnetic tension force at 20 Mm altitude increased ({absolute value}) from {-}$2.5 \times10^{-8}$ to {-}$7 \times10^{-8}$ N/m$^2$ after the emergence of the magnetic bipole, and it subsequently decreased from {-}$7 \times10^{-8}$ to {-}$5 \times10^{-8}$ N/m$^2$ in the magnetic cancellation phase. Additionally, we calculated the decay index of the external potential magnetic field of the filament using the formula $n = -d\ln|B|/d\ln|h|$~\citep{kliem06}, where $B$ represents the external potential magnetic field strength and $h$ denotes the height above the solar surface. The calculated magnetic decay indexes are plotted in \nfig{fig4}(h). One can see that the emergence of the magnetic bipole resulted in a significant increase of the magnetic decay index above the filament below the altitude of about 100 Mm (see the blue and green curves in \nfig{fig4}(h)). During the late cancellation phase at 02:12 UT on January 5, the decay index had recovered to the level before the flux emergence. Previous statistical and theoretical studies have established that the critical value ($n_{ct}$) for torus instability is typically in the range of 1.1--1.3~\citep[e.g.,][]{2010ApJ...718.1388D}. In our case, the decay index profiles increased with height. If we take 1.2 as the critical value (the red dashed line in \nfig{fig4}(h)), the critical altitudes for torus instability before and after the emergence of the magnetic bipole are about 58 and 45 Mm, respectively.

During the interaction and reconnection between the emerging magnetic bipole and the filament, we noted the interesting dynamic length change of the thin filament threads within the yellow box region in \nfig{fig1} (a) (supplementary animation online). To further investigate this phenomenon, we generated a time-distance diagram along the filament threads (red curve S3 in \nfig{fig5}). The diagram reveals a gradual increase in the filament thread length over time with a growth rate of approximately 117 m~s$^{-1}$ (see the region between the two red dashed lines in \nfig{fig5} (a)). To quantify the filament thread length, we selected three representative time points for a detailed measurement based on the AIA 171~\AA~images at 08:19 UT on January 3 (pre-emergence phase), 20:08 UT on January 3 (late emergence phase), and 01:58 UT on January 5 (late cancellation phase). The filament thread lengths were derived from the Gaussian full width at half maximum (FWHM) of intensity profiles along cuts 1--3 (see \nfig{fig5} (b)--(g)). Before the bipole's emergence, the filament thread length was about 5.6 Mm. After the emergence (also the {breaking} of the filament{-magnetic field}), the filament thread length grew to 10 Mm, increased by about 80\% compared to the original state before the emergence. The filament threads further grew to 21.3 Mm during the flux cancellation phase, increased by about 113\% compared to the time point right after the emergence. This dynamic change of the filament thread length is consistent with the simulation results presented by \cite{2012ApJ...746...30L,2021ApJ...920..131G,2016ApJ...817..157L}. The authors found that in stable flux ropes, the length of the filament threads exhibits a linear relationship with the curvature radius of the magnetic dips; a larger curvature radius of the magnetic dips is associated with longer filament threads. It should be noted that the filament threads also show some oscillations (see \nfig{fig5} (a)). However, this phenomenon is beyond the aim of our study, we will discuss this intriguing observation in our future work.

\section{Physical Factors suppressing the Filament Eruption}\label{sec:interp}
The growth of the filament threads suggests an increase in the radius of curvature of the filament’s magnetic dip. As the interaction and reconnection between the filament and the emerging magnetic bipole progressed, the dip became progressively flatter, exhibiting a larger radius of curvature~\citep{2012ApJ...748L..26X, 2003ApJ...593.1187K, 2021ApJ...920..131G,2016ApJ...817..157L,2012ApJ...746...30L}. Following the reconnection-unfavored scenario described in \citet{chen00}, the emergence of a bipole with {a reconnection-unfavorable direction within the filament channel} induces a downward motion of the flux rope. This motion is primarily driven by the enhanced magnetic tension force exerted by the confined field of the filament. A similar enhancement in downward magnetic tension force was evidenced in our case by investigating the extrapolated coronal magnetic fields overlying the filament at different stages, suggesting that the studied filament likely underwent a height reduction, akin to the scenario proposed by Chen. In our case, as shown in the green curve in \nfig{fig4} (h), the decay index at 10–30 Mm heights where the flux rope typically resides, first increases then decreases when a long filament breaks. The filament channel’s stabilization is likely due to the decay index returning to values below the critical threshold for torus instability.

As is well known, for a toroidal filament flux rope, {the stability of a filament channel} can also be explained by the force balance between the outward hoop force generated by the toroidal current and the internal poloidal magnetic field and the inward strapping force generated by the toroidal current and the external poloidal magnetic field \citep{2015Natur.528..526M}. Due to observational limitations, we focus only on the upward hoop force in this study.

   \begin{equation}
    F_p = \frac{\pi I_t^2}{c^2} \left[ \ln \left( \frac{8R}{a} \right) - 1 \right]
    \label{eq:hoop_force}
  \end{equation}
 
\begin{itemize}
    \item $I_t$:  Toroidal current for a toroidal flux rope (unit: $\mathrm{A}$)
    \item $R$: Major radius of the  current loop (unit: $\mathrm{length}$)
    \item $a$: Minor radius of the current loop (unit: $\mathrm{length}$)
\end{itemize}

Another possible explanation for the increased curvature radius of the magnetic dip exists: the filament's magnetic field may have reduced its twist number during the processes of magnetic flux emergence and cancellation. This could result from the injection of positive twist from emerging bipole into the dextral-chirality filament channel through magnetic reconnection. Such helicity transfer would decrease the twist number of the filament’s magnetic field and its toroidal current $I_t$, thereby weakening the hoop force of the core filament and lowering its height due to the reduced upward net force. This process would further contribute to an eruption-unfavored condition for the filament. Overall, the {flux emergence} in our event suppressed filament eruption through three primary mechanisms:
(1) Enhancing the downward magnetic tension force of the confined field.
(2) Reducing the upward net force of the core filament.
(3) The decreased decay index at the height of the filament flux rope.

The subsequent magnetic reconnection between SF1 and SF2 recovered the configuration of the original filament channel, which further extended the filament threads to a length of 21.3 Mm. This means that the radius of curvature of the magnetic dip structure increased significantly, and the magnetic dip became flatter, which corresponded to the reduced magnetic tension ($\mathbf{F}_{\text{tension}} = \frac{B^2\kappa}{\mu_0}$) of the filament dip structure~\citep{2022A&A...660A..54L}, where $\kappa$ denotes the curvature of a magnetic dip structure.

\begin{figure*}
\centering
\includegraphics[width=0.95\textwidth]{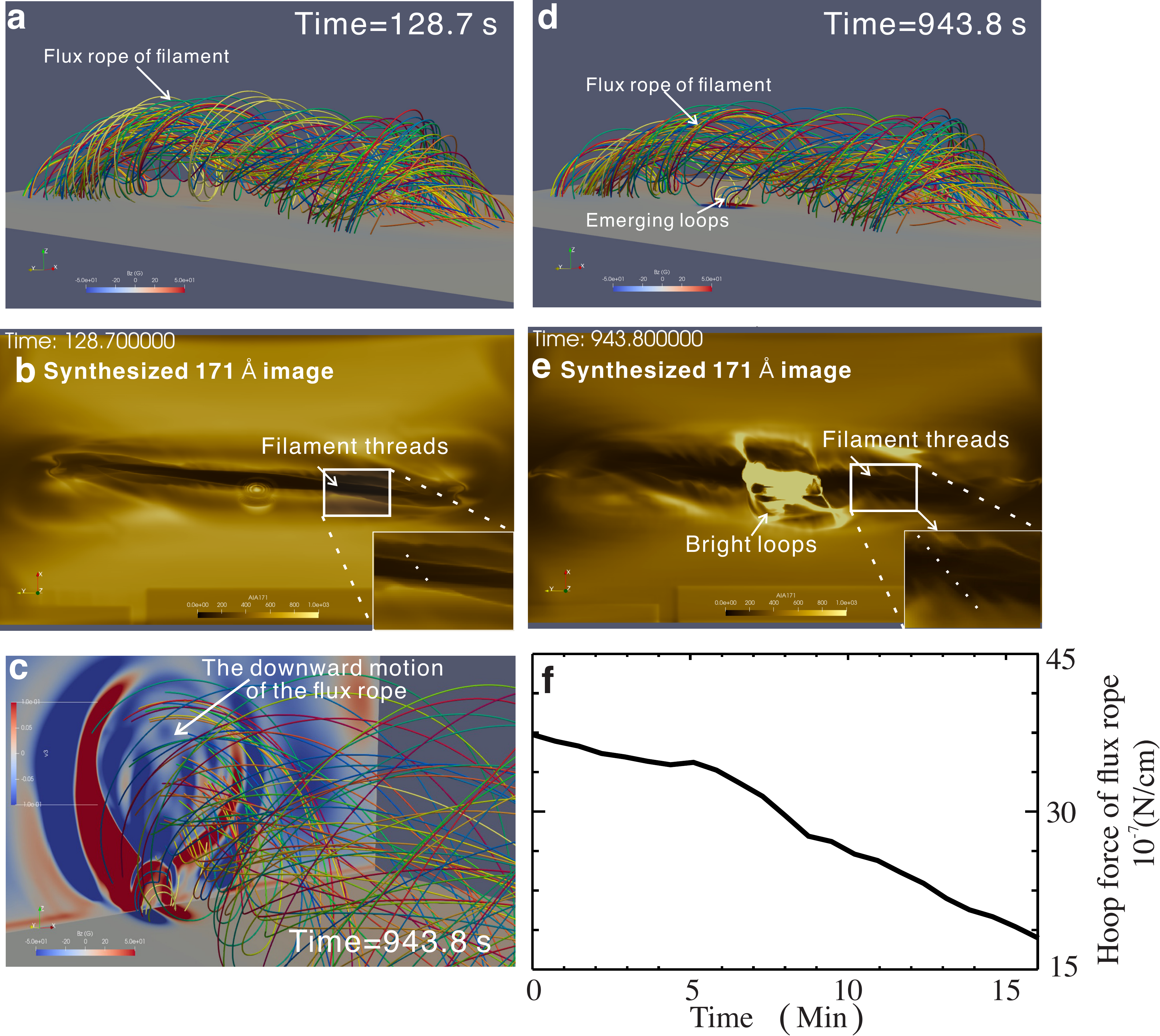}
\caption{Panels (a) and (d) show the magnetic flux rope of the simulated filament at 128.7 seconds and 943.8 seconds. Panels (b) and (e) show the simultaneous AIA 171 \AA\ synthetic images; the filament threads labeled by the white dotted lines. Panel (c) shows the downward motions of the filament, and panel (f) shows the curve of hoop force of the flux rope. {An animation of this simulation is available in the online Journal. The animation has two panels: the evolution of the magnetic flux rope and the synthetic AIA 171 \AA\ images. The animation corresponds to $t={0}--{943}$ seconds of the simulation.}
\label{fig6}}
\end{figure*}

\section{Numerical modeling}\label{sec:modeling}

To model a fully ionized and magnetized plasma, we employed the semi-relativistic magnetohydrodynamic (MHD) equations, with the energy solution being the internal energy equation. These equations combine the pre-Maxwell equations of electromagnetism and the equations of fluid dynamics as follows:

\begin{equation}
\frac{\partial\rho}{\partial t} + \nabla\cdot (\rho\mathbf{v})= 0
\end{equation}

\begin{equation}
\begin{split}
\frac{\partial\left(\rho\mathbf{v} + \frac{\mathbf{E}\times\mathbf{B}}{c^{2}\mu_0}\right)}{\partial t}
+ \nabla\cdot \Biggl[ \rho\mathbf{v} \otimes \mathbf{v} 
+ \left(p + \frac{\mathbf{B}^{2}}{2\mu_0}\right)\mathbf{I} \\
- \frac{\mathbf{B} \otimes \mathbf{B}}{\mu_0} 
+ \frac{\mathbf{E}^{2}\mathbf{I}}{2c^{2}\mu_0} 
- \frac{\mathbf{E} \otimes \mathbf{E}}{c^{2}\mu_0} \Biggr]\\ 
= \frac{1}{\mu_0}\left(\frac{1}{c_0^{2}} - \frac{1}{c^{2}}\right)\mathbf{E}\nabla \cdot \mathbf{E} + \rho\mathbf{g},
\end{split}
\end{equation}

\begin{equation}
\frac{\partial\left(\frac{p}{\gamma - 1}\right)}{\partial t} 
+ \nabla\cdot \left(\frac{p}{\gamma - 1}\mathbf{v}\right) 
= -p\nabla\cdot \mathbf{v} + \mathcal{H} - \mathcal{R} 
+ \nabla\cdot (\kappa \nabla T), \label{eq:3} 
\end{equation}

\begin{equation}
\frac{\partial\mathbf{B}}{\partial t} 
+ \nabla\cdot (\mathbf{v} \otimes \mathbf{B} - \mathbf{B} \otimes \mathbf{v}) 
= \mathbf{0}
 \label{eq:4} 
 \end{equation}
 
\begin{equation}
\nabla \cdot \mathbf{B} = 0. 
\label{eq:5}
\end{equation}

We conducted MHD simulations in the low-$\beta$ regime of the emergence of a twisted magnetic flux tube into a pre-existing prominence magnetic structure constructed by the Titov-Démoulin-modified (TDm) model~\citep{2014ApJ...790..163T}. The flux rope of the prominence is embedded in an idealized bipolar background field $B_q$, which is modeled by two fictitious point sources of strength $q = 500$ G at a depth of 200 Mm below the photospheric boundary and at a distance of 20 Mm from each other. The integral axis path of the flux ropes follows a circular-arc shape in the vertical plane of the configuration. The great-circle distance of the arc is 113.731 Mm, the shortest-circle distance of the arc is 30 Mm, and the torus minor radius is set to $a = 30$ Mm.

To lift a twisted flux tube into the domain, our simulation adopted an approximation based on \citet{2003ApJ...589L.105F}. We drove the computational domain at the lower electric field boundary using $V\times B$. The twisted flux tube is derived from \citet{2012A&A...537A..62A}, with a floating speed of v = \speed{1.75}, and the depth of the twisted flux rope beneath the photosphere is 27 Mm. For this work, we chose $B_T=-60$ G, $R = 20$ Mm, $r = 4$ Mm, and $\alpha = 16$.

We used the open-source MPI-AMRVAC code \citep{2018ApJS..234...30X, 2023A&A...673A..66K} to solve the semi-relativistic MHD equations in a conservative form with a finite-volume scheme. In our setup, we employed an HLL solver with a Cada3-type limiter for reconstruction and a four-step Runge-Kutta method for time integration.

As in the 3D MHD simulation study of \cite{2016ApJ...823...22X}, the density of the prominence is lightly formed in the concave-up dips. We attached the prominence density to the concave-up dips in the prominence flux rope structure. \nfig{fig6} (a) shows the magnetic flux rope structure of the filament at 128.7 seconds. The following equations determine the dips and density of the prominence:

\begin{equation}
dip=\nabla B_z\cdot\mathbf{\hat{B}}
\end{equation}

\begin{equation}
\begin{split}
\rho_{prom}=2.341668\cdot10^{-15}\cdot\left(12-0.3\cdot z\right) g/cm^{3}\\
 \left(5 Mm \le z \le 20 Mm, dip \ge 0, \frac{B_z}{B} \le 0.008\right) \\
\end{split}
\end{equation}

The initial background heating is given by

\begin{equation}
\bm{\hat{b}} = \frac{\bm{\hat{B}}}{\|\bm{\hat{B}}\|}
\label{eq:unit_b}
\end{equation}

\begin{equation}
\bm{\kappa} = \bm{\hat{b}} \cdot \nabla \bm{\hat{b}}
\label{eq:kappa_def}
\end{equation}

\begin{equation}
\alpha = 
\begin{cases}
0.1, & \text{if } \|\kappa\| < 0.1 \\
1.0, & \text{if } \|\kappa\| > 1.0 \\
\|\kappa\|, & \text{otherwise}
\end{cases}
\label{eq:kappa_limit}
\end{equation}

\begin{equation}
\begin{split}
	\mathcal{H}_0 = 10^{-4}\left(0.5\times B^{1.56} \alpha^{0.75}\rho^{0.125}+0.5\times\exp\left( -\frac{z}{20} \right) \right)\text{erg cm}^{-3}.
\end{split}
	\label{eq:bh}
\end{equation}

To model the global coronal heating mechanism sustaining a hot corona, we adopt a mixed heating framework proposed by~\citet{2022A&A...668A..47B} (see Equation~\ref{eq:bh}). One term in this formulation follows the magnetic curvature-dependent prescriptions of~\citet{2013ApJ...773..134L} and~\citet{2016ApJ...817...15M}, where $\kappa$ denotes the local magnetic field curvature. Accounting for energy transport from lower to upper atmospheric layers, we hypothesize that heating efficiency diminishes with increasing altitude, as motivated by previous studies~\citep{1981ApJ...243..288S,2002ApJS..142..269A}. This leads to the definition of an additional term as a multivariate power-law function, which satisfied heating efficiency attenuates with increasing altitude.

In our simulations, we computed the hoop force at the central interface of the flux rope. Our MHD simulations successfully reproduced the key observational features of the filament channel evolution during magnetic emergence. Specifically, the simulations captured: (1) the elongation of filament threads (see the white dotted lines in panels (b) and (e)). (2) the downward motions of the core filament (see \nfig{fig6} (c)). (3) the suppression mechanism of filament eruption through decreased upward hoop force (see \nfig{fig6} (f)). Crucially, the elongation of filament threads during the magnetic flux emergence phase also reveals the magnetic dip flattening and flux rope apex lowering (see smovie1.mp4). These findings confirm that the {flux} emergence can suppress filament eruptions through reducing the upward net force of the filament. 

\section{Conclusions and Discussions}\label{sec:summary}

Using multi-wavelength observations from the SDO, we investigated a rare case of magnetic flux emergence beneath a quiescent solar filament and its role in suppressing filament eruptions, trying to answer the question of how magnetic flux emergence suppresses the eruption of solar filaments. The emerging bipole beneath the filament interacted with its horizontal magnetic field through sequential reconnection events, generating recurrent two-sided coronal jets while maintaining overall stability. The suppression mechanism operated through three interconnected processes: First, the emergence enhanced downward magnetic tension forces in the confining field while simultaneously reducing the upward hoop force of the filament's flux rope. Second, the decay index of the filament channel returned to a lower value (\( < 1.2 \)) at the height of the flux rope, below the critical threshold for torus instability. Third, the twist transfer from the positive-twist emerging flux to the dextral-chirality filament likely reduced the core flux rope's twist, further stabilizing the system. The elongation of filament threads during magnetic flux emergence and cancellation phases revealed how magnetic dip flattening and flux rope apex lowering modified the eruption thresholds, demonstrating the importance of topological evolution caused by flux emergence and cancellation in the suppression of filament eruptions. These findings challenge simplistic applications of decay index criteria and establish that flux emergence can actively suppress eruptions through combined force rebalancing and topological modification. Future studies should quantify twist transfer efficiency in such scenarios and investigate how these suppression mechanisms scale across different solar activity regimes.

To date, the two-sided jets model has been developed through multiple observational studies \citep{1999SoPh..190..167A, 2013ApJ...775..132J, shen19a, yangbo19, 2024ApJ...964....7Y,2024ApJ...973...74K,2025arXiv250402396T,yangbo19,2017ApJ...851...67S}, {and is strongly supported by theoretical simulations of 3D magnetic reconnection topologies~\citep[e.g.,][]{2008ApJ...673L.211M,2013ApJ...771...20M,2018ApJ...864..165W}. These simulations demonstrate that bidirectional outflows are an inherent feature of such configurations, where the orientation of the current sheet dictates whether one direction (e.g., upflow in coronal hole jets;~\citet{2019ApJ...873...93K,tang21}) dominates or flows remain symmetric.} Recent high spatiotemporal resolution observations suggest that {the eruptive magnetic structure of} the mini-filaments may drive two-sided jets. For example, \citet{shen19a} documented two-sided jets resulting from magnetic reconnections between {the eruptive magnetic structure of} the mini-filaments and pre-existing horizontal filament fields. Earlier, \citet{1995Natur.375...42Y} proposed an alternative triggering mechanism involving transient emerging coronal loops reconnecting with pre-existing horizontal fields. However, clear observational verification of two-sided loop jets based on this model remains limited.  In this study, with the emerging bipole located beneath the filament, magnetic reconnection occurred between the filament's horizontal magnetic field lines and the emerging loops, generating two-sided jets. Consequently, our observations support the model proposing that two-sided jets arise from magnetic reconnection between newly emerging coronal loops of a magnetic bipole and the pre-existing horizontal coronal magnetic structures of a filament.

In the study of previous filament eruptions, torus instability---an important filament eruption mechanism---typically occurs in filament eruption events~\citep[e.g.,][]{Tor05,kliem06, Liuy08, Liu12}. {In this study, we employ the torus instability model to investigate the impact of magnetic emergence within the filament channel.} This instability emerges when considering the filament as a flux-rope system. Previous research demonstrated that the formation of torus instability in a flux rope magnetic system depends on the decay index of the external magnetic field, with theoretical calculations revealing a critical value threshold $n_{ct}$ for the filament flux rope system. The flux rope remains stable only when $n \textless n_{ct}$; otherwise, it will rapidly erupt due to torus instability~\citep{1978mit..book.....B,kliem06,2010ApJ...718.1388D}. The critical value has been extensively studied for decades: \citet{1978mit..book.....B} initially derived $n_{ct}=1.5$ for a thin circular current, while \citet{kliem06} further confirmed this value in the limit of very large aspect ratios. Conversely, \citet{2010ApJ...718.1388D} showed that the critical value for flux ropes ranges between 1.1--1.3 for both circular and straight current channels. In our observed event, the filament's decay index ($n$) reached 1.2 at 45 Mm during magnetic emergence, exceeding conventional instability thresholds ($n_{cr}=1.1-1.3$). Surprisingly, the filament channel remained stable, with its stabilization attributed to the decay index returning to a lower value that lies below the critical threshold for torus instability.

The emerging bipolar case we identified closely resembles the reconnection-unfavored scenario described by \citet{chen00}, wherein unfavourable magnetic emergence augments the magnetic tension of the overlying arcade field, thereby inhibiting filament eruption. This mechanism precisely corresponds to our observation that the emerging bipole increased magnetic tension force by 38\%, effectively suppressing the filament eruption. Moreover, magnetic reconnection between the newly emerged coronal loops of the bipole and the filament's horizontal magnetic field broke the filament channel into two segments. This finding aligns with \citet{2018ApJ...862..117D}'s research, which utilized SDO multi-wavelength observations and three-dimensional MHD simulations to demonstrate how a newly emerging positive twist bipolar region can segment a filament channel without triggering its eruption. Our observations revealed that the filament's breaking significantly enlarged the curvature radius of magnetic dips, as manifested by the elongation of filament threads. These morphological changes indicated a reduced hoop force and an elevated critical value of the filament channel, substantiating that flux emergence suppresses filament eruptions through three distinct mechanisms:
(1) The decay index got back to a lower value, remained below critical value for torus instability,
(2) Enhancing the downward magnetic tension force of the overlying arcade field,
(3) Diminishing the outward hoop force generated by the filament's toroidal current.

Measuring the coronal magnetic field of filaments remains a significant challenge, prompting most studies to rely on indirect techniques such as tracking filament fine-structure evolution via jets, nonlinear force-free field extrapolation, and filament oscillation analysis \citep{2014ApJ...786..151S,2014ApJ...795..130S,2022MNRAS.516L..12T,2024ApJ...964..125S,2021ApJ...923...45Z,2025arXiv250414984Z}. As critical fine-structural elements, filament threads serve as essential tracers of the underlying magnetic configuration \citep{2003ApJ...593.1187K, 2021ApJ...920..131G, 2024ApJ...965..160C}, and in this study, we systematically documented their length variations across three phases—pre-emergence (5.6 Mm), late emergence (10 Mm), and late cancellation (21.3 Mm)—a progressive elongation reflecting topological transformations in the filament’s magnetic structure. {To investigate whether photospheric shearing influenced this evolution, we analyzed vector magnetic field data from DAVE4VM; however, coherent flow measurements along the quiescent filament channel were hindered by low signal-to-noise ratios in quiet-Sun regions, yielding a velocity field with highly turbulent characteristics and no organized shear patterns. While our MHD simulations (Section 5) demonstrate that flux emergence-driven magnetic reconnection alone can reproduce the observed thread elongation and dip flattening without requiring photospheric shear, we acknowledge that unresolved convective motions or weak shear along the channel could still contribute to such structural changes.} However, these empirical observations and simulation align with theoretical frameworks linking filament thread formation to magnetic topology \citep{2003ApJ...593.1187K, 2021ApJ...920..131G, 2015ApJ...814L..17S, 2024ApJ...965..160C}, providing direct evidence for the interplay between fine-scale threads and large-scale magnetic evolution. Thus, we propose that filament threads represent critical observational constraints for characterizing the complex coronal magnetic fields supporting solar filaments, bridging challenges in direct magnetic field measurement with fine-structural dynamics.

\begin{acknowledgments}
The authors are grateful for the excellent data provided by the SDO teams. The authors gratefully acknowledge the constructive comments and insightful suggestions provided by the reviewers, which have significantly improved the quality of this work. This work is supported by the Natural Science Foundation of China (12173083), Shenzhen Key Laboratory Launching Project (No. ZDSYS20210702140800001) and the Specialized Research Fund for State Key Laboratory of Solar Activity and Space Weather, the Strategic Priority Research Program of the Chinese Academy of Sciences (XDB0560000) and the Key Research Program of Frontier Sciences, CAS (grant No. ZDBS-LY-SLH013). The numerical computations were conducted on the Yunnan University Astronomy Supercomputer.
\end{acknowledgments}


\begin{thebibliography}{}
\expandafter\ifx\csname natexlab\endcsname\relax\def\natexlab#1{#1}\fi
\providecommand{\url}[1]{\href{#1}{#1}}
\providecommand{\dodoi}[1]{doi:~\href{http://doi.org/#1}{\nolinkurl{#1}}}
\providecommand{\doeprint}[1]{\href{http://ascl.net/#1}{\nolinkurl{http://ascl.net/#1}}}
\providecommand{\doarXiv}[1]{\href{https://arxiv.org/abs/#1}{\nolinkurl{https://arxiv.org/abs/#1}}}

\bibitem[{{Alexander} \& {Fletcher}(1999)}]{1999SoPh..190..167A}
{Alexander}, D., \& {Fletcher}, L. 1999, \solphys, 190, 167,
  \dodoi{10.1023/A:1005213826793}

\bibitem[{{Archontis} \& {Hood}(2012)}]{2012A&A...537A..62A}
{Archontis}, V., \& {Hood}, A.~W. 2012, \aap, 537, A62,
  \dodoi{10.1051/0004-6361/201116956}

\bibitem[{{Archontis} {et~al.}(2004){Archontis}, {Moreno-Insertis},
  {Galsgaard}, {Hood}, \& {O'Shea}}]{2004A&A...426.1047A}
{Archontis}, V., {Moreno-Insertis}, F., {Galsgaard}, K., {Hood}, A., \&
  {O'Shea}, E. 2004, \aap, 426, 1047, \dodoi{10.1051/0004-6361:20035934}

\bibitem[{{Aschwanden} \& {Schrijver}(2002)}]{2002ApJS..142..269A}
{Aschwanden}, M.~J., \& {Schrijver}, C.~J. 2002, \apjs, 142, 269,
  \dodoi{10.1086/341945}

\bibitem[{{Bateman}(1978)}]{1978mit..book.....B}
{Bateman}, G. 1978, {MHD instabilities}

\bibitem[{{Berger} \& {Field}(1984)}]{1984JFM...147..133B}
{Berger}, M.~A., \& {Field}, G.~B. 1984, Journal of Fluid Mechanics, 147, 133,
  \dodoi{10.1017/S0022112084002019}

\bibitem[{{Bose} {et~al.}(2024){Bose}, {De Pontieu}, {Hansteen}, {Sainz Dalda},
  {Savage}, \& {Winebarger}}]{2024NatAs...8..697B}
{Bose}, S., {De Pontieu}, B., {Hansteen}, V., {et~al.} 2024, Nature Astronomy,
  8, 697, \dodoi{10.1038/s41550-024-02241-8}

\bibitem[{{Brughmans} {et~al.}(2022){Brughmans}, {Jenkins}, \&
  {Keppens}}]{2022A&A...668A..47B}
{Brughmans}, N., {Jenkins}, J.~M., \& {Keppens}, R. 2022, \aap, 668, A47,
  \dodoi{10.1051/0004-6361/202244071}

\bibitem[{{Chen} {et~al.}(2024){Chen}, {Xia}, \& {Chen}}]{2024ApJ...965..160C}
{Chen}, H., {Xia}, C., \& {Chen}, H. 2024, \apj, 965, 160,
  \dodoi{10.3847/1538-4357/ad3352}

\bibitem[{{Chen} {et~al.}(2016){Chen}, {Zhang}, {Li}, \&
  {Ma}}]{2016ApJ...818L..27C}
{Chen}, H., {Zhang}, J., {Li}, L., \& {Ma}, S. 2016, \apjl, 818, L27,
  \dodoi{10.3847/2041-8205/818/2/L27}

\bibitem[{{Chen}(2011)}]{2011LRSP....8....1C}
{Chen}, P.~F. 2011, Living Reviews in Solar Physics, 8, 1,
  \dodoi{10.12942/lrsp-2011-1}

\bibitem[{{Chen} \& {Shibata}(2000)}]{chen00}
{Chen}, P.~F., \& {Shibata}, K. 2000, \apj, 545, 524, \dodoi{10.1086/317803}

\bibitem[{{Chen} {et~al.}(2020){Chen}, {Xu}, \& {Ding}}]{2020RAA....20..166C}
{Chen}, P.-F., {Xu}, A.-A., \& {Ding}, M.-D. 2020, Research in Astronomy and
  Astrophysics, 20, 166, \dodoi{10.1088/1674-4527/20/10/166}

\bibitem[{{Dacie} {et~al.}(2018){Dacie}, {T{\"o}r{\"o}k}, {D{\'e}moulin},
  {Linton}, {Downs}, {van Driel-Gesztelyi}, {Long}, \&
  {Leake}}]{2018ApJ...862..117D}
{Dacie}, S., {T{\"o}r{\"o}k}, T., {D{\'e}moulin}, P., {et~al.} 2018, \apj, 862,
  117, \dodoi{10.3847/1538-4357/aacce3}

\bibitem[{{Dalmasse} {et~al.}(2014){Dalmasse}, {Pariat}, {D{\'e}moulin}, \&
  {Aulanier}}]{2014SoPh..289..107D}
{Dalmasse}, K., {Pariat}, E., {D{\'e}moulin}, P., \& {Aulanier}, G. 2014,
  \solphys, 289, 107, \dodoi{10.1007/s11207-013-0326-4}

\bibitem[{{D{\'e}moulin} \& {Aulanier}(2010)}]{2010ApJ...718.1388D}
{D{\'e}moulin}, P., \& {Aulanier}, G. 2010, \apj, 718, 1388,
  \dodoi{10.1088/0004-637X/718/2/1388}

\bibitem[{{D{\'e}moulin} \& {Berger}(2003)}]{2003SoPh..215..203D}
{D{\'e}moulin}, P., \& {Berger}, M.~A. 2003, \solphys, 215, 203,
  \dodoi{10.1023/A:1025679813955}

\bibitem[{{Dubey} {et~al.}(2006){Dubey}, {van der Holst}, \&
  {Poedts}}]{2006A&A...459..927D}
{Dubey}, G., {van der Holst}, B., \& {Poedts}, S. 2006, \aap, 459, 927,
  \dodoi{10.1051/0004-6361:20054719}

\bibitem[{{Fan} \& {Gibson}(2003)}]{2003ApJ...589L.105F}
{Fan}, Y., \& {Gibson}, S.~E. 2003, \apjl, 589, L105, \dodoi{10.1086/375834}

\bibitem[{{Feynman} \& {Martin}(1995)}]{1995JGR...100.3355F}
{Feynman}, J., \& {Martin}, S.~F. 1995, \jgr, 100, 3355,
  \dodoi{10.1029/94JA02591}

\bibitem[{{Guo} {et~al.}(2021){Guo}, {Zhou}, {Guo}, {Ni}, {Karpen}, \&
  {Chen}}]{2021ApJ...920..131G}
{Guo}, J.~H., {Zhou}, Y.~H., {Guo}, Y., {et~al.} 2021, \apj, 920, 131,
  \dodoi{10.3847/1538-4357/ac17e8}

\bibitem[{{Harden} {et~al.}(2021){Harden}, {Panesar}, {Moore}, {Sterling}, \&
  {Adams}}]{2021ApJ...912...97H}
{Harden}, A.~R., {Panesar}, N.~K., {Moore}, R.~L., {Sterling}, A.~C., \&
  {Adams}, M.~L. 2021, \apj, 912, 97, \dodoi{10.3847/1538-4357/abee19}

\bibitem[{{Hou} {et~al.}(2020){Hou}, {Li}, {Song}, \& {Zhang}}]{hou20}
{Hou}, Y.~J., {Li}, T., {Song}, Z.~P., \& {Zhang}, J. 2020, \aap, 640, A101,
  \dodoi{10.1051/0004-6361/202038348}

\bibitem[{{Ji} {et~al.}(2003){Ji}, {Wang}, {Schmahl}, {Moon}, \&
  {Jiang}}]{ji03}
{Ji}, H., {Wang}, H., {Schmahl}, E.~J., {Moon}, Y.~J., \& {Jiang}, Y. 2003,
  \apjl, 595, L135, \dodoi{10.1086/378178}

\bibitem[{{Jiang} {et~al.}(2013){Jiang}, {Bi}, {Yang}, {Li}, {Yang}, \&
  {Zheng}}]{2013ApJ...775..132J}
{Jiang}, Y., {Bi}, Y., {Yang}, J., {et~al.} 2013, \apj, 775, 132,
  \dodoi{10.1088/0004-637X/775/2/132}

\bibitem[{{Jiang} {et~al.}(2011){Jiang}, {Yang}, {Hong}, {Bi}, \&
  {Zheng}}]{jiang11}
{Jiang}, Y., {Yang}, J., {Hong}, J., {Bi}, Y., \& {Zheng}, R. 2011, \apj, 738,
  179, \dodoi{10.1088/0004-637X/738/2/179}

\bibitem[{{Joshi} {et~al.}(2014){Joshi}, {Srivastava}, {Filippov}, {Kayshap},
  {Uddin}, {Chandra}, {Prasad Choudhary}, \& {Dwivedi}}]{joshi14}
{Joshi}, N.~C., {Srivastava}, A.~K., {Filippov}, B., {et~al.} 2014, \apj, 787,
  11, \dodoi{10.1088/0004-637X/787/1/11}

\bibitem[{{Joshi} {et~al.}(2013){Joshi}, {Srivastava}, {Filippov}, {Uddin},
  {Kayshap}, \& {Chandra}}]{joshi13}
---. 2013, \apj, 771, 65, \dodoi{10.1088/0004-637X/771/1/65}

\bibitem[{{Karpen} {et~al.}(2003){Karpen}, {Antiochos}, {Klimchuk}, \&
  {MacNeice}}]{2003ApJ...593.1187K}
{Karpen}, J.~T., {Antiochos}, S.~K., {Klimchuk}, J.~A., \& {MacNeice}, P.~J.
  2003, \apj, 593, 1187, \dodoi{10.1086/376690}

\bibitem[{{Keppens} {et~al.}(2023){Keppens}, {Popescu Braileanu}, {Zhou},
  {Ruan}, {Xia}, {Guo}, {Claes}, \& {Bacchini}}]{2023A&A...673A..66K}
{Keppens}, R., {Popescu Braileanu}, B., {Zhou}, Y., {et~al.} 2023, \aap, 673,
  A66, \dodoi{10.1051/0004-6361/202245359}

\bibitem[{{Kliem} \& {T{\"o}r{\"o}k}(2006)}]{kliem06}
{Kliem}, B., \& {T{\"o}r{\"o}k}, T. 2006, \prl, 96, 255002,
  \dodoi{10.1103/PhysRevLett.96.255002}

\bibitem[{{Kumar} {et~al.}(2019){Kumar}, {Karpen}, {Antiochos}, {Wyper},
  {DeVore}, \& {DeForest}}]{2019ApJ...873...93K}
{Kumar}, P., {Karpen}, J.~T., {Antiochos}, S.~K., {et~al.} 2019, \apj, 873, 93,
  \dodoi{10.3847/1538-4357/ab04af}

\bibitem[{{Kumar} {et~al.}(2024){Kumar}, {Karpen}, {Yurchyshyn}, {DeVore}, \&
  {Antiochos}}]{2024ApJ...973...74K}
{Kumar}, P., {Karpen}, J.~T., {Yurchyshyn}, V., {DeVore}, C.~R., \&
  {Antiochos}, S.~K. 2024, \apj, 973, 74, \dodoi{10.3847/1538-4357/ad63a2}

\bibitem[{{Kumar} {et~al.}(2011){Kumar}, {Srivastava}, {Filippov},
  {Erd{\'e}lyi}, \& {Uddin}}]{kumar11}
{Kumar}, P., {Srivastava}, A.~K., {Filippov}, B., {Erd{\'e}lyi}, R., \&
  {Uddin}, W. 2011, \solphys, 272, 301, \dodoi{10.1007/s11207-011-9829-z}

\bibitem[{{Lemen} {et~al.}(2012){Lemen}, {Title}, {Akin}, {Boerner}, {Chou},
  {Drake}, {Duncan}, {Edwards}, {Friedlaender}, {Heyman}, {Hurlburt}, {Katz},
  {Kushner}, {Levay}, {Lindgren}, {Mathur}, {McFeaters}, {Mitchell}, {Rehse},
  {Schrijver}, {Springer}, {Stern}, {Tarbell}, {Wuelser}, {Wolfson}, {Yanari},
  {Bookbinder}, {Cheimets}, {Caldwell}, {Deluca}, {Gates}, {Golub}, {Park},
  {Podgorski}, {Bush}, {Scherrer}, {Gummin}, {Smith}, {Auker}, {Jerram},
  {Pool}, {Soufli}, {Windt}, {Beardsley}, {Clapp}, {Lang}, \&
  {Waltham}}]{lemen12}
{Lemen}, J.~R., {Title}, A.~M., {Akin}, D.~J., {et~al.} 2012, \solphys, 275,
  17, \dodoi{10.1007/s11207-011-9776-8}

\bibitem[{{Li} {et~al.}(2025){Li}, {Zhang}, {Ying}, {Feng}, {Su}, {Lin}, \&
  {Zhang}}]{2025RAA....25a5006L}
{Li}, S.-Y., {Zhang}, Q.-M., {Ying}, B.-L., {et~al.} 2025, Research in
  Astronomy and Astrophysics, 25, 015006, \dodoi{10.1088/1674-4527/ad9a33}

\bibitem[{{Lionello} {et~al.}(2013){Lionello}, {Winebarger}, {Mok}, {Linker},
  \& {Miki{\'c}}}]{2013ApJ...773..134L}
{Lionello}, R., {Winebarger}, A.~R., {Mok}, Y., {Linker}, J.~A., \&
  {Miki{\'c}}, Z. 2013, \apj, 773, 134, \dodoi{10.1088/0004-637X/773/2/134}

\bibitem[{{Liu} {et~al.}(2012){Liu}, {Wang}, {Shen}, \& {Wang}}]{Liu12}
{Liu}, K., {Wang}, Y., {Shen}, C., \& {Wang}, S. 2012, \apj, 744, 168,
  \dodoi{10.1088/0004-637X/744/2/168}

\bibitem[{{Liu}(2008)}]{Liuy08}
{Liu}, Y. 2008, \apjl, 679, L151, \dodoi{10.1086/589282}

\bibitem[{{Liu} \& {Schuck}(2012)}]{2012ApJ...761..105L}
{Liu}, Y., \& {Schuck}, P.~W. 2012, \apj, 761, 105,
  \dodoi{10.1088/0004-637X/761/2/105}

\bibitem[{{Lu} {et~al.}(2024){Lu}, {Chen}, {Ding}, {Wang}, {Dai}, \&
  {Cheng}}]{2024NatAs...8..706L}
{Lu}, Z., {Chen}, F., {Ding}, M.~D., {et~al.} 2024, Nature Astronomy, 8, 706,
  \dodoi{10.1038/s41550-024-02244-5}

\bibitem[{{Luna} {et~al.}(2012){Luna}, {Karpen}, \&
  {DeVore}}]{2012ApJ...746...30L}
{Luna}, M., {Karpen}, J.~T., \& {DeVore}, C.~R. 2012, \apj, 746, 30,
  \dodoi{10.1088/0004-637X/746/1/30}

\bibitem[{{Luna} {et~al.}(2022){Luna}, {Terradas}, {Karpen}, \&
  {Ballester}}]{2022A&A...660A..54L}
{Luna}, M., {Terradas}, J., {Karpen}, J., \& {Ballester}, J.~L. 2022, \aap,
  660, A54, \dodoi{10.1051/0004-6361/202142907}

\bibitem[{{Luna} {et~al.}(2016){Luna}, {Terradas}, {Khomenko}, {Collados}, \&
  {de Vicente}}]{2016ApJ...817..157L}
{Luna}, M., {Terradas}, J., {Khomenko}, E., {Collados}, M., \& {de Vicente}, A.
  2016, \apj, 817, 157, \dodoi{10.3847/0004-637X/817/2/157}

\bibitem[{{Lynch} \& {Edmondson}(2013)}]{Lynch13}
{Lynch}, B.~J., \& {Edmondson}, J.~K. 2013, \apj, 764, 87,
  \dodoi{10.1088/0004-637X/764/1/87}

\bibitem[{{Martin}(1998)}]{1998SoPh..182..107M}
{Martin}, S.~F. 1998, \solphys, 182, 107, \dodoi{10.1023/A:1005026814076}

\bibitem[{{Mok} {et~al.}(2016){Mok}, {Miki{\'c}}, {Lionello}, {Downs}, \&
  {Linker}}]{2016ApJ...817...15M}
{Mok}, Y., {Miki{\'c}}, Z., {Lionello}, R., {Downs}, C., \& {Linker}, J.~A.
  2016, \apj, 817, 15, \dodoi{10.3847/0004-637X/817/1/15}

\bibitem[{{Moreno-Insertis} \& {Galsgaard}(2013)}]{2013ApJ...771...20M}
{Moreno-Insertis}, F., \& {Galsgaard}, K. 2013, \apj, 771, 20,
  \dodoi{10.1088/0004-637X/771/1/20}

\bibitem[{{Moreno-Insertis} {et~al.}(2008){Moreno-Insertis}, {Galsgaard}, \&
  {Ugarte-Urra}}]{2008ApJ...673L.211M}
{Moreno-Insertis}, F., {Galsgaard}, K., \& {Ugarte-Urra}, I. 2008, \apjl, 673,
  L211, \dodoi{10.1086/527560}

\bibitem[{{Morgan} \& {Druckm{\"u}ller}(2014)}]{2014SoPh..289.2945M}
{Morgan}, H., \& {Druckm{\"u}ller}, M. 2014, \solphys, 289, 2945,
  \dodoi{10.1007/s11207-014-0523-9}

\bibitem[{{Myers} {et~al.}(2015){Myers}, {Yamada}, {Ji}, {Yoo}, {Fox},
  {Jara-Almonte}, {Savcheva}, \& {Deluca}}]{2015Natur.528..526M}
{Myers}, C.~E., {Yamada}, M., {Ji}, H., {et~al.} 2015, \nat, 528, 526,
  \dodoi{10.1038/nature16188}

\bibitem[{{Pariat} {et~al.}(2005){Pariat}, {D{\'e}moulin}, \&
  {Berger}}]{2005A&A...439.1191P}
{Pariat}, E., {D{\'e}moulin}, P., \& {Berger}, M.~A. 2005, \aap, 439, 1191,
  \dodoi{10.1051/0004-6361:20052663}

\bibitem[{{Parker}(1988)}]{1988ApJ...330..474P}
{Parker}, E.~N. 1988, \apj, 330, 474, \dodoi{10.1086/166485}

\bibitem[{{Pesnell} {et~al.}(2012){Pesnell}, {Thompson}, \&
  {Chamberlin}}]{pesnell12}
{Pesnell}, W.~D., {Thompson}, B.~J., \& {Chamberlin}, P.~C. 2012, \solphys,
  275, 3, \dodoi{10.1007/s11207-011-9841-3}

\bibitem[{{Schou} {et~al.}(2012){Schou}, {Scherrer}, {Bush}, {Wachter},
  {Couvidat}, {Rabello-Soares}, {Bogart}, {Hoeksema}, {Liu}, {Duvall}, {Akin},
  {Allard}, {Miles}, {Rairden}, {Shine}, {Tarbell}, {Title}, {Wolfson},
  {Elmore}, {Norton}, \& {Tomczyk}}]{Schou2012}
{Schou}, J., {Scherrer}, P.~H., {Bush}, R.~I., {et~al.} 2012, \solphys, 275,
  229, \dodoi{10.1007/s11207-011-9842-2}

\bibitem[{{Schuck}(2008)}]{2008ApJ...683.1134S}
{Schuck}, P.~W. 2008, \apj, 683, 1134, \dodoi{10.1086/589434}

\bibitem[{{Serio} {et~al.}(1981){Serio}, {Peres}, {Vaiana}, {Golub}, \&
  {Rosner}}]{1981ApJ...243..288S}
{Serio}, S., {Peres}, G., {Vaiana}, G.~S., {Golub}, L., \& {Rosner}, R. 1981,
  \apj, 243, 288, \dodoi{10.1086/158597}

\bibitem[{{Shen} {et~al.}(2014{\natexlab{a}}){Shen}, {Ichimoto}, {Ishii},
  {Tian}, {Zhao}, \& {Shibata}}]{2014ApJ...786..151S}
{Shen}, Y., {Ichimoto}, K., {Ishii}, T.~T., {et~al.} 2014{\natexlab{a}}, \apj,
  786, 151, \dodoi{10.1088/0004-637X/786/2/151}

\bibitem[{{Shen} {et~al.}(2015){Shen}, {Liu}, {Liu}, {Chen}, {Su}, {Xu}, \&
  {Liu}}]{2015ApJ...814L..17S}
{Shen}, Y., {Liu}, Y., {Liu}, Y.~D., {et~al.} 2015, \apjl, 814, L17,
  \dodoi{10.1088/2041-8205/814/1/L17}

\bibitem[{{Shen} {et~al.}(2012{\natexlab{a}}){Shen}, {Liu}, \& {Su}}]{shen12b}
{Shen}, Y., {Liu}, Y., \& {Su}, J. 2012{\natexlab{a}}, \apj, 750, 12,
  \dodoi{10.1088/0004-637X/750/1/12}

\bibitem[{{Shen} {et~al.}(2012{\natexlab{b}}){Shen}, {Liu}, {Su}, \&
  {Deng}}]{shen12a}
{Shen}, Y., {Liu}, Y., {Su}, J., \& {Deng}, Y. 2012{\natexlab{b}}, \apj, 745,
  164, \dodoi{10.1088/0004-637X/745/2/164}

\bibitem[{{Shen} {et~al.}(2017{\natexlab{a}}){Shen}, {Liu}, {Tian}, \&
  {Qu}}]{2017ApJ...851..101S}
{Shen}, Y., {Liu}, Y., {Tian}, Z., \& {Qu}, Z. 2017{\natexlab{a}}, \apj, 851,
  101, \dodoi{10.3847/1538-4357/aa9af0}

\bibitem[{{Shen} {et~al.}(2014{\natexlab{b}}){Shen}, {Liu}, {Chen}, \&
  {Ichimoto}}]{2014ApJ...795..130S}
{Shen}, Y., {Liu}, Y.~D., {Chen}, P.~F., \& {Ichimoto}, K. 2014{\natexlab{b}},
  \apj, 795, 130, \dodoi{10.1088/0004-637X/795/2/130}

\bibitem[{{Shen} {et~al.}(2017{\natexlab{b}}){Shen}, {Liu}, {Su}, {Qu}, \&
  {Tian}}]{2017ApJ...851...67S}
{Shen}, Y., {Liu}, Y.~D., {Su}, J., {Qu}, Z., \& {Tian}, Z. 2017{\natexlab{b}},
  \apj, 851, 67, \dodoi{10.3847/1538-4357/aa9a48}

\bibitem[{{Shen} {et~al.}(2019{\natexlab{a}}){Shen}, {Qu}, {Zhou}, {Duan},
  {Tang}, \& {Yuan}}]{shen19b}
{Shen}, Y., {Qu}, Z., {Zhou}, C., {et~al.} 2019{\natexlab{a}}, \apjl, 885, L11,
  \dodoi{10.3847/2041-8213/ab4cf3}

\bibitem[{{Shen} {et~al.}(2019{\natexlab{b}}){Shen}, {Qu}, {Yuan}, {Chen},
  {Duan}, {Zhou}, {Tang}, {Huang}, \& {Liu}}]{shen19a}
{Shen}, Y., {Qu}, Z., {Yuan}, D., {et~al.} 2019{\natexlab{b}}, \apj, 883, 104,
  \dodoi{10.3847/1538-4357/ab3a4d}

\bibitem[{{Shen} {et~al.}(2024){Shen}, {Liu}, {Yao}, {Zhou}, {Tang}, {Qu},
  {Zhou}, {Duan}, {Tan}, \& {Ibrahim}}]{2024ApJ...964..125S}
{Shen}, Y., {Liu}, D., {Yao}, S., {et~al.} 2024, \apj, 964, 125,
  \dodoi{10.3847/1538-4357/ad2349}

\bibitem[{{Song} {et~al.}(2020){Song}, {Hou}, {Zhang}, \&
  {Wang}}]{2020ApJ...892...79S}
{Song}, Z., {Hou}, Y., {Zhang}, J., \& {Wang}, P. 2020, \apj, 892, 79,
  \dodoi{10.3847/1538-4357/ab77b3}

\bibitem[{{Tan} {et~al.}(2022){Tan}, {Shen}, {Zhou}, {Duan}, {Tang}, {Zhou}, \&
  {Yao}}]{2022MNRAS.516L..12T}
{Tan}, S., {Shen}, Y., {Zhou}, X., {et~al.} 2022, \mnras, 516, L12,
  \dodoi{10.1093/mnrasl/slac069}

\bibitem[{{Tang} {et~al.}(2025){Tang}, {Shen}, {Zhou}, {Yao}, \&
  {Liu}}]{2025arXiv250402396T}
{Tang}, Z., {Shen}, Y., {Zhou}, C., {Yao}, S., \& {Liu}, D. 2025, arXiv
  e-prints, arXiv:2504.02396, \dodoi{10.48550/arXiv.2504.02396}

\bibitem[{{Tang} {et~al.}(2021){Tang}, {Shen}, {Zhou}, {Duan}, {Zhou}, {Tan},
  \& {Elmhamdi}}]{tang21}
{Tang}, Z., {Shen}, Y., {Zhou}, X., {et~al.} 2021, \apjl, 912, L15,
  \dodoi{10.3847/2041-8213/abf73a}

\bibitem[{{Tian} {et~al.}(2017){Tian}, {Liu}, {Shen}, {Elmhamdi}, {Su}, {Liu},
  \& {Kordi}}]{tian2017}
{Tian}, Z., {Liu}, Y., {Shen}, Y., {et~al.} 2017, \apj, 845, 94,
  \dodoi{10.3847/1538-4357/aa8095}

\bibitem[{{Titov} {et~al.}(2014){Titov}, {T{\"o}r{\"o}k}, {Mikic}, \&
  {Linker}}]{2014ApJ...790..163T}
{Titov}, V.~S., {T{\"o}r{\"o}k}, T., {Mikic}, Z., \& {Linker}, J.~A. 2014,
  \apj, 790, 163, \dodoi{10.1088/0004-637X/790/2/163}

\bibitem[{{T{\"o}r{\"o}k} \& {Kliem}(2005)}]{Tor05}
{T{\"o}r{\"o}k}, T., \& {Kliem}, B. 2005, \apjl, 630, L97,
  \dodoi{10.1086/462412}

\bibitem[{{T{\"o}r{\"o}k} {et~al.}(2024){T{\"o}r{\"o}k}, {Linton}, {Leake},
  {Miki{\'c}}, {Lionello}, {Titov}, \& {Downs}}]{2024ApJ...962..149T}
{T{\"o}r{\"o}k}, T., {Linton}, M.~G., {Leake}, J.~E., {et~al.} 2024, \apj, 962,
  149, \dodoi{10.3847/1538-4357/ad1826}

\bibitem[{{T{\"o}r{\"o}k} {et~al.}(2011){T{\"o}r{\"o}k}, {Panasenco}, {Titov},
  {Miki{\'c}}, {Reeves}, {Velli}, {Linker}, \& {De Toma}}]{tor11}
{T{\"o}r{\"o}k}, T., {Panasenco}, O., {Titov}, V.~S., {et~al.} 2011, \apjl,
  739, L63, \dodoi{10.1088/2041-8205/739/2/L63}

\bibitem[{{Wang} \& {Ji}(2013)}]{2013ScChD..56.1091W}
{Wang}, J., \& {Ji}, H. 2013, Science China Earth Sciences, 56, 1091,
  \dodoi{10.1007/s11430-013-4648-8}

\bibitem[{{Wyper} {et~al.}(2018){Wyper}, {DeVore}, {Karpen}, {Antiochos}, \&
  {Yeates}}]{2018ApJ...864..165W}
{Wyper}, P.~F., {DeVore}, C.~R., {Karpen}, J.~T., {Antiochos}, S.~K., \&
  {Yeates}, A.~R. 2018, \apj, 864, 165, \dodoi{10.3847/1538-4357/aad9f7}

\bibitem[{{Xia} {et~al.}(2012){Xia}, {Chen}, \&
  {Keppens}}]{2012ApJ...748L..26X}
{Xia}, C., {Chen}, P.~F., \& {Keppens}, R. 2012, \apjl, 748, L26,
  \dodoi{10.1088/2041-8205/748/2/L26}

\bibitem[{{Xia} \& {Keppens}(2016)}]{2016ApJ...823...22X}
{Xia}, C., \& {Keppens}, R. 2016, \apj, 823, 22,
  \dodoi{10.3847/0004-637X/823/1/22}

\bibitem[{{Xia} {et~al.}(2018){Xia}, {Teunissen}, {El Mellah}, {Chan{\'e}}, \&
  {Keppens}}]{2018ApJS..234...30X}
{Xia}, C., {Teunissen}, J., {El Mellah}, I., {Chan{\'e}}, E., \& {Keppens}, R.
  2018, \apjs, 234, 30, \dodoi{10.3847/1538-4365/aaa6c8}

\bibitem[{{Xu} {et~al.}(2020){Xu}, {Su}, {Chen}, {Ruan}, {Awasthi}, {Zhang},
  {Zhang}, {Ji}, {Zhang}, \& {Liu}}]{xuh20}
{Xu}, H., {Su}, J., {Chen}, J., {et~al.} 2020, \apj, 901, 121,
  \dodoi{10.3847/1538-4357/abb01d}

\bibitem[{{Xue} {et~al.}(2017){Xue}, {Yan}, {Yang}, {Wang}, \&
  {Zhao}}]{2017ApJ...840L..23X}
{Xue}, Z., {Yan}, X., {Yang}, L., {Wang}, J., \& {Zhao}, L. 2017, \apjl, 840,
  L23, \dodoi{10.3847/2041-8213/aa7066}

\bibitem[{{Yang} {et~al.}(2016){Yang}, {Jiang}, {Yang}, {Yu}, \&
  {Xu}}]{2016ApJ...816...41Y}
{Yang}, B., {Jiang}, Y., {Yang}, J., {Yu}, S., \& {Xu}, Z. 2016, \apj, 816, 41,
  \dodoi{10.3847/0004-637X/816/1/41}

\bibitem[{{Yang} {et~al.}(2019){Yang}, {Yang}, {Bi}, {Xu}, {Hong}, {Li}, \&
  {Chen}}]{yangbo19}
{Yang}, B., {Yang}, J., {Bi}, Y., {et~al.} 2019, \apj, 887, 220,
  \dodoi{10.3847/1538-4357/ab557e}

\bibitem[{{Yang} {et~al.}(2024){Yang}, {Chen}, {Hong}, {Yang}, \&
  {Bi}}]{2024ApJ...964....7Y}
{Yang}, J., {Chen}, H., {Hong}, J., {Yang}, B., \& {Bi}, Y. 2024, \apj, 964, 7,
  \dodoi{10.3847/1538-4357/ad23e5}

\bibitem[{{Yang} {et~al.}(2012){Yang}, {Jiang}, {Zheng}, {Bi}, {Hong}, \&
  {Yang}}]{yang12}
{Yang}, J., {Jiang}, Y., {Zheng}, R., {et~al.} 2012, \apj, 745, 9,
  \dodoi{10.1088/0004-637X/745/1/9}

\bibitem[{{Yang} {et~al.}(2023){Yang}, {Yan}, {Xue}, {Wang}, {Yang}, {Li},
  {Xu}, {Peng}, {Sun}, \& {Zhang}}]{2023ApJ...943...62Y}
{Yang}, L., {Yan}, X., {Xue}, Z., {et~al.} 2023, \apj, 943, 62,
  \dodoi{10.3847/1538-4357/aca9d2}

\bibitem[{{Yao} {et~al.}(2024){Yao}, {Shen}, {Zhou}, {Liu}, \&
  {Zhou}}]{2024ApJ...975L...5Y}
{Yao}, S., {Shen}, Y., {Zhou}, C., {Liu}, D., \& {Zhou}, X. 2024, \apjl, 975,
  L5, \dodoi{10.3847/2041-8213/ad84ea}

\bibitem[{{Yokoyama} \& {Shibata}(1995)}]{1995Natur.375...42Y}
{Yokoyama}, T., \& {Shibata}, K. 1995, \nat, 375, 42, \dodoi{10.1038/375042a0}

\bibitem[{{Zheng} {et~al.}(2017){Zheng}, {Zhang}, {Chen}, {Wang}, {Du}, {Li},
  \& {Yang}}]{zhengr17}
{Zheng}, R., {Zhang}, Q., {Chen}, Y., {et~al.} 2017, \apj, 836, 160,
  \dodoi{10.3847/1538-4357/aa5c38}

\bibitem[{{Zhou} {et~al.}(2025){Zhou}, {Shen}, {Xia}, {Liu}, {Tang}, \&
  {Yao}}]{2025arXiv250414984Z}
{Zhou}, C., {Shen}, Y., {Xia}, C., {et~al.} 2025, arXiv e-prints,
  arXiv:2504.14984, \dodoi{10.48550/arXiv.2504.14984}

\bibitem[{{Zhou} {et~al.}(2021){Zhou}, {Shen}, {Zhou}, {Tang}, {Duan}, \&
  {Tan}}]{2021ApJ...923...45Z}
{Zhou}, C., {Shen}, Y., {Zhou}, X., {et~al.} 2021, \apj, 923, 45,
  \dodoi{10.3847/1538-4357/ac28a0}

\end{thebibliography}
\end{document}